\documentclass[eprint,amsmath,amssymb,aps, superscriptaddress]{revtex4-2}
\usepackage[utf8]{inputenc}
\usepackage{graphicx}
\usepackage{graphics}
\usepackage{dcolumn}
\usepackage{bm}
\usepackage{amsmath}
\usepackage{amssymb}
\usepackage{mathtools}
\usepackage{tikz,siunitx}
\usepackage{float}
\usepackage{verbatim}
\usepackage{booktabs}
\usepackage{xcolor}
\usepackage{url}
\usepackage[pdfencoding=auto]{hyperref}
\usepackage[colorinlistoftodos]{todonotes}
\usepackage[normalem]{ulem}

\newcommand{\eqn}[1]{\begin{align}#1\end{align}}
\newcommand{\bs}[1]{\boldsymbol{#1}}

\newcommand{\pare}[1]{\left( #1 \right) }
\newcommand{\corchete}[1]{\left[ #1 \right]}
\newcommand{\fr}[2]{\frac{#1}{#2}}
\newcommand{\wtil}[1]{\widetilde{#1}}
\newcommand{\mc}[1]{\mathcal{#1}}
\newcommand{\avg}[1]{\langle #1 \rangle}
\newcommand{\tex}[1]{\mbox{\scriptsize{#1}}}

\definecolor{darkgreen}{rgb}{0, 0.5, 0.05}

\definecolor{darkgreen}{rgb}{0, 0.5, 0.05}
\newcommand{\itodo}[1]{\todo[inline,backgroundcolor=blue!02]{#1}}

\newcommand{\deleted}[1]{}

\def\dd{\mathrm{d}}  
\def\kt{k_B T}
\def\bna{\bs{\nabla}}
\def\bDl{\Delta \bs{l}}

\def\bC{\bs{C}}

\def\bD{\bs{D}}

\def\bx{\bs{x}}
\def\bI{\bs{I}}

\def\bK{\bs{K}}

\def\bM{\bs{M}}
\def\bR{\bs{R}}
\def\bomega{\bs{\omega}}
\def\bW{\bs{W}}

\def\btau{\bs \tau}
\def\bn{\bs{n}}

\def\bB{\bs{B}}

\def\bu{\bs{u}}
\def\bv{\bs{v}}

\def\bU{\bs{U}}
\def\bl{\bs l}
\def\blambda{\bs{\lambda}}
\def\br{\bs{r}}
\def\bF{\bs{F}}
\def\bbf{\bs{f}}
\def\bphi{\bs{\phi}}
\def\bq{\bs{q}}

\def\bg{\bs{g}}

\def\bN{\bs{N}}

\def\by{\bs{y}}

\def\bZ{\bs{Z}}

\def\bzero{\bs{0}}
\def\btheta{\bs \theta}

\def\bsigma{\bs \sigma}

\def\mcB{\mc{B}}
\def\bmG{\bs{\mc{G}}}
\def\bmK{\bs{\mc{K}}}

\def\bmZ{\bs{\mc{Z}}}

\begin{document}


\title{Modeling complex particle suspensions: perspectives on the rigid multiblob method}

\author{Blaise Delmotte}
\email{blaise.delmotte@cnrs.fr}
 \affiliation{LadHyX, CNRS, Ecole Polytechnique, Institut Polytechnique de Paris, 91120 Palaiseau, France}

\author{Florencio Balboa Usabiaga}
\email{fbalboa@bcamath.org}
 \affiliation{BCAM - Basque Center for Applied Mathematics, Mazarredo 14, Bilbao, E48009, Basque Country - Spain}

\begin{abstract}

Many suspensions contain particles with complex shapes that are affected not only by hydrodynamics, but also by thermal fluctuations, internal activity, kinematic constraints and other long-range non-hydrodynamic interactions.
Modeling these systems represents a significant numerical challenge due to the interplay between different effects
and the need to accurately capture multiscale phenomena.
In this article we review recent developments to model large suspensions of particles of arbitrary shapes and multiple
couplings with controllable accuracy within the rigid multiblob framework.
We discuss the governing equations, highlight key numerical developments, and illustrate applications ranging from microswimmers to complex colloidal suspensions.
This review illustrates the effectiveness and versatility of the rigid multiblob method in tackling a wide range of physical problems in fluid mechanics, soft matter physics, biophysics, materials and colloidal science. 
  
\end{abstract}

                    
\maketitle

\tableofcontents

\section{Introduction}

\begin{figure}[t!]
  \centering
  \includegraphics[width=0.75\textwidth]{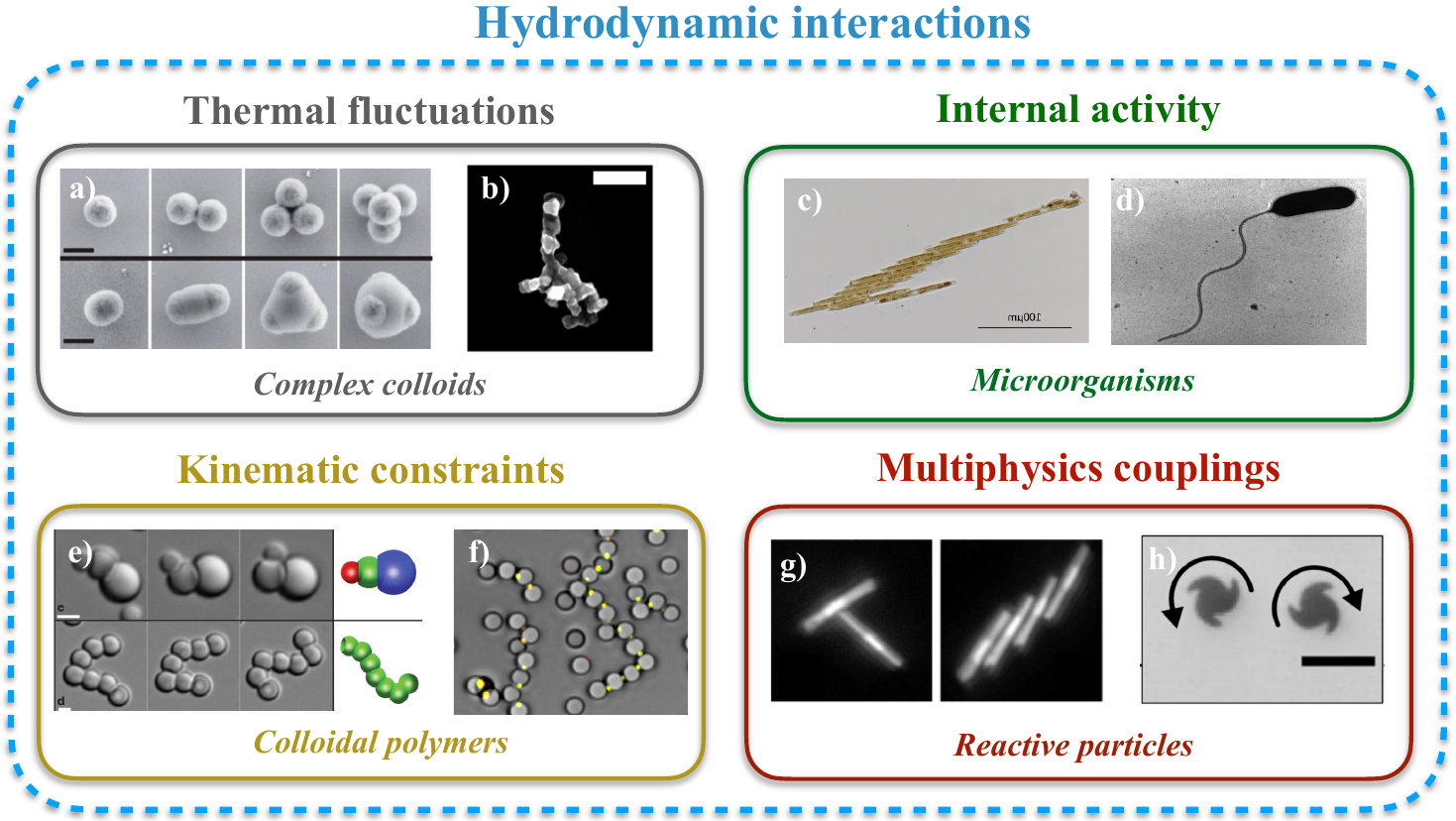}
  \caption{Illustration of various effects at play in complex suspensions.  
  Each box represents one of these effects and shows  experimental illustrations.  a) Patchy colloids used to form gels or suspensions with complex structures \cite{li2020colloidal}. Scale bars: $1 \si{\micro m}$. b) Fumed silica particle used to tune the rheology of colloidal suspensions \cite{bourrianne2022tuning}. Scale bar: $150 \si{\nano  m}$. c) Diatom chain \textit{Bacillaria Paxillifer} made of sliding rod-like cells \cite{dskeet}. Scale bar: $100 \si{\micro m}$. d) Bacterium \textit{Vibrio Cholerae} swimming with its flagellum \cite{syed2010vibrio}. e) Lock and key colloids \cite{Sacanna2010}. Scale bars: $2 \si{\micro m}$. f) Chains of freely-jointed droplets \cite{McMullen2018}. g) Self assembly of catalytic rods \cite{Wykes2016}. h) Spontaneous rotation of chiral catalytic particles \cite{Brooks2019}. Scale bar: $20 \si{\micro m}$.
  }
  \label{fig:context}
\end{figure}

Suspensions of small, micron-sized,  particles  are ubiquitous in biological, environmental and industrial systems. 
Their typical size and speed are such that, at their scale, the Reynolds number is very small (Re $\ll1$) and the surrounding flow follows the Stokes equations. 
In this limit, the hydrodynamic interactions between the particles are long-ranged, meaning that all particle motions are instantaneously coupled through the fluid.

In many applications, suspended particles have complex shapes and are subject to additional effects than hydrodynamic interactions,  such as thermal fluctuations (i.e.\ Brownian motion), internal activity, kinematic constraints, or non-hydrodynamic long-ranged interactions (e.g.\ chemical, magnetic, electrostatic...), see Fig.\ \ref{fig:context}. These effects, in combination with the particle shape,  lead to complex dynamics at the scale of the particles, such as non-trivial diffusion, self-assembly, rotation, or propulsion,  and provide intricate macroscopic properties at the scale of the suspensions, such as complex rheological and elastic responses, collective diffusion, collective motion, flow generation and mixing.






The interplay between the individual  behavior of complex particles and their  suspension dynamics has attracted significant attention in the past decades.
Experimental methods for synthesizing complex suspensions and numerical tools for modeling them have developed rapidly.
However, significant hurdles remain in the numerical domain. 
Indeed, the modeling of complex suspensions poses multiple challenges due to particle geometry, thermal fluctuations with multiplicative noise, nontrivial boundary conditions, kinematic constraints, active deformations or multiphysics couplings (see Fig.\ref{fig:context}).
In addition, these effects must be accurately captured to bridge the gap between the particle scale and the macroscopic suspension scale.
Traditional numerical methods, such as  Boundary Element Methods,  are accurate and can now handle large collections of spherical \cite{Yan2020b}, or even spheroidal particles \cite{crowder2025boundary}, but they remain restricted to few particles of complex shape due to their high computational cost
and cannot handle thermal fluctuations (see Ref.\ \cite{bao2018fluctuating} for an exception in 2D).

An alternative approach is the rigid multiblob framework. In the rigid multiblob  (RMB) method, also  called composite bead model \cite{Swan2016}, or Stokesian Dynamics of rigid assemblies \cite{BradyBossis1988}, rigid bodies are discretized using a collection of minimally resolved spherical markers, also called \emph{blobs}, or beads, which are constrained to move as a rigid body.  This idea  first emerged nearly 50 years ago \cite{mccammon1976frictional} to compute the friction coefficients of non-spherical aggregates. Unlike point-forces, i.e.\ Stokeslets, which are singular, these blobs have a finite size which allows to compute their hydrodynamic interactions, and thus the hydrodynamic coefficients of the aggregates, with regularized Green's functions, i.e.\ mobility matrices, which were first derived by Rotne $\&$ Prager \cite{Rotne1969} and Yamakawa \cite{Yamakawa1970} in the early 70s. This seminal multiblob idea was further formalized and developed in the late 1970s and early 1980s \cite{nakajima1977general,de1977hydrodynamic,de1977hydrodynamicsrot}, see \cite{de1981hydrodynamic} for a  review of these pioneering works.
Nowadays, the multiblob method is still widely used  to compute hydrodynamic coefficients  (mobility, diffusion and resistance matrices) of rigid bodies such as colloids, molecules or clusters \cite{zurita2007motion,Bringley2008,pandey2016flow,cichocki1995stokes,fernandes2002brownian,de2000calculation,lobaskin2004new,kutteh2010rigid,dlugosz2013hydrodynamic,poblete2014hydrodynamics,ortega2011prediction,gastaldi2011distribution,molina2013direct,Reichert2005,Cortez2005,Codutti2018,Bianchi2020,Swan2016}. 
However, since the 1990s, and more particularly in the past decade, several groups across the world have extensively developed the method  to extend its domains of application  beyond the motion of passive  colloidal clusters, but also to  improve its numerical performances, using  efficient summation techniques, modern linear algebra algorithms and grid-based numerical methods  \cite{Hinsen1995,Swan2011,Delong2015,Swan2016,Fiore2019,Usabiaga2016,Sprinkle2017,Usabiaga2022,Westwood2022,Delmotte2024,Gidituri2024, krucker2024immersed}. 

While it shares some similarities with the well-known method of Regularized Stokeslets  \cite{Cortez2001,Cortez2005}, in which bodies are discretized with regularized point forces, the rigid multiblob method additionally handles thermal fluctuations and effectively couples with grid-based methods (such as Immersed Boundary methods \cite{Peskin2002,Maxey2001}) to solve hydrodynamic interactions between blobs, whereas the former can only be used for deterministic problems and relies exclusively on analytical Green's functions.


In this Perspective, intended for a wide audience of fluid dynamicists and soft matter physicists, we present recent developments of the rigid multiblob method  to model suspensions containing a large number of particles of arbitrary shapes and subject to multiple physical couplings, with controllable accuracy and high scalability.
These recent efforts open new perspectives for the modeling of complex particle suspensions and their intricate phenomena. 
We first introduce the governing equations of complex particle suspensions in Section \ref{sec: governing}, with an emphasis on the modeling challenges that they represent. Then we introduce the rigid multiblob framework in Section \ref{sec:RMM} and present recent developments that successfully address these modeling challenges.
Finally we showcase some examples and applications where the method has been used in Section \ref{sec:applications} and outline future directions regarding the development of the method and its use to address new physical phenomena in Section \ref{sec:perspectives}.

The framework presented in this article is implemented in a code which is free software, user friendly for non-experts and continuously enriched by a group  of collaborators
in a GitHub repository (\url{https://github.com/stochasticHydroTools/RigidMultiblobsWall}). The repository contains many hands-on examples that users can run on a basic laptop. 
This code has been actively used in the community, in combination with experiments and theory, to study a variety of systems involving (micro-)swimmers in various geometries \citep{Dombrowski2019,Usabiaga2022,Gidituri2024a,Gidituri2024,Gidituri2024a}, passive, active and reactive   colloids
\citep{driscoll2017unstable, Sprinkle2020,Sprinkle2021,Brosseau2019,Brosseau2021,Van2023, Delmotte2024,chen2024restructuring},  diffusing  molecules \citep{Dlugosz2022,Dlugosz2023}, suspensions of sedimenting cubes  \cite{Kundu2025}, viruses \citep{Moreno2022}, functionalized nanoparticles \citep{Moreno2023}, M\"obius strips  \citep{Moreno2024},  branched particles suspensions \citep{Westwood2022,Balboa2024}, fibers in structured environments \citep{Makanga2023,Makanga2024}, coated drops \citep{Gao2023},  magnetic spheres \citep{Bililign2022},  chains \citep{Yang2020} or sheets  \citep{Wang2023}.


\section{Governing equations and modeling challenges}
\label{sec: governing}

This section introduces the equations that govern the motion of complex suspensions in viscous flows. Complexity is introduced at multiple levels of the system: in the shape of the particles, in  their surface velocity and deformation, in the fluid through thermal fluctuations,  in the relative motion between particles due to  kinematic constraints,  or with non-hydrodynamic couplings such as chemical interactions. We first introduce the governing equations of the hydrodynamic problem and then extend them to  include thermal fluctuations. Then we  explain how the equations of motion are modified in the presence of kinematic constraints. 
Finally, we tackle the situation where particles interact through additional physical couplings and provide the    governing equations for the case of chemical and  phoretic interactions between the suspended particles. 
At each step, we briefly discuss the main modeling challenges that will be addressed in  Section \ref{sec:RMM}.

\subsection{Governing equations for the hydrodynamic problem}
\label{sec:governing_HI}
Let $\{\mcB_p\}_{p=1}^M$ be a set of $M$  bodies immersed in a Stokes flow.
The configuration of each  body $p$ is described by the location of a tracking point, $\bq_p$,
and its orientation represented by the unit quaternion $\btheta_p$, or in compact notation $\bx_p=\{\bq_p, \btheta_p\}$.
The linear and angular velocities of the tracking point are $\bu_p$ and $\bomega_p$.
The external force and torque applied to a body are $\bbf_p$ and $\btau_p$.
We will use the concise notation $\bU_p=\{\bu_p, \bomega_p\}$ and $\bF_p=\{\bbf_p, \btau_p\}$ as well, while
 unscripted vectors will refer to the composite vector formed by the variables of all the bodies,
e.g.\ $\bU = \{\bU_p\}_{p=1}^M$. 
The velocity and pressure, $\bv$ and $p$, of the fluid with viscosity $\eta$ are governed by the Stokes equations  \citep{Kim1991}
\eqn{
  \label{eq:Stokes}
  -\bna p + \eta \bna^2 \bv &= \bzero, \\
  \label{eq:div}
  \bna \cdot \bv &= 0,
}
while at the bodies surface the fluid obeys the boundary condition 
\eqn{
  \label{eq:no-slip-continuum}
  \bv(\br) = (\bs{\mc{K}} \bU_p)(\br) + \bu_s(\br) = \bu_p + \bomega_p \times (\br - \bq_p) + \bu_s(\br)\; \mbox{for }\br \in \partial \mcB_p,
}
where we have introduced the geometric operator $\bs{\mc{K}}$ of rigid body motion. $\bu_s$ is the surface velocity of the particle relative to its rigid body motion. This velocity can correspond to the displacement of the surface due to active deformations, also called swimming gait \cite{Swan2011}, it can also  be interpreted as the velocity of an ambient flow (e.g.\ shear, vortical or quadratic flows), or as an active slip velocity due  to  phoresis  \citep{Anderson1989,Golestanian2007} or ciliary motion \citep{Blake1971a}  evaluated at the surface of a rigid particle.
Boundary conditions also apply at the bounding surface of the domain considered.

Since inertia does not play a role in Stokes flows the conservation of linear and angular momentum
reduces to the balance of force and torque. For every body $p$ the balance between hydrodynamic and external stresses is given by 
\eqn{
  \label{eq:balanceF_continuum}
  \int_{\partial \mcB_p} \blambda(\br) \,\dd S_{\br} = \bbf_p, \\
  \label{eq:balanceT_continuum}
  \int_{\partial \mcB_p} (\br - \bq_p) \times \blambda(\br) \,\dd S_{\br} = \btau_p,  
}
where $\blambda(\br)$ is a constraint force per unit area  that enforces the  boundary condition \eqref{eq:no-slip-continuum}.
In absence of surface velocity ($\bu_s=\bzero$), $-\blambda$ is the hydrodynamic traction exerted on the bodies by the fluid,
i.e.\ $-\blambda(\br) = \bsigma\cdot\bn$ where $\bsigma = -p\bI + 2\eta(\bna\bv + \bna\bv^T)$ is the fluid stress tensor \cite{Pozrikidis1992, Smith2021}.
The adjoint of the geometric operator $\bs{\mc{K}}$ can be used to write the balance of force and torque for all bodies as
\footnote{
The adjoint property ensures that the power dissipated by the motion of the particle surface integrated over the whole surface matches the power dissipated by the motion of
a rigid particle.
For a particle without surface velocity ($\bu_s=\bzero$)
\eqn{
  P &= \int_{\partial \mcB_p} \blambda^T(\br) \, \bv(\br) \; \dd S_{\br} = \int_{\partial \mcB_p} \blambda^T(\br)\, (\bmK \bU)(\br) \; \dd S_{\br} = \bU^T \bmK^T \blambda = \bU^T \bF
  = \bu \cdot \bbf + \btau \cdot \bomega.
}}
\eqn{
\bs{\mc{K}}^T \blambda = \bF.
}

In order to compute the trajectories of the immersed bodies, the equations of motion are integrated in time
  \eqn{
    \fr{\dd \bx}{\dd t} = \bU.
    \label{eq:dx_dt}
  }
  The integration of unit quaternions requires some care to correctly represent rotations, details are discussed in \citep{Delong2015, Westwood2022}.
  
  If the particle surface moves relative to the rigid body motion, then the surface shape can be integrated in time using \eqref{eq:no-slip-continuum}.

\subsubsection{The mobility problem}
Given the nonhydrodynamic forces and torques acting on the rigid bodies, $\bF$, and  their surface velocity $\bu_s$, the equations \eqref{eq:Stokes}-\eqref{eq:balanceT_continuum} can be solved
for the bodies velocities $\bU$ and the constraint forces $\blambda$. Since the Stokes equations are linear we can write the velocity of $M$ rigid bodies as
\eqn{
  \label{eq:U_NF}
  \bU = \bN \bF - \wtil{\bN}\bu_s,
}
where $\bN=\bN(\bx)$ is the $6M \times 6M$ \textit{body mobility matrix} that couple the forces and torques acting on the rigid bodies with their velocities, and $\wtil{\bN}$ is a linear operator ($\wtil{\bN}:L^2(\partial \mcB,\mathbb{R}^3)\rightarrow \mathbb{R}^{6M}$)\footnote{
$L^2(\partial \mcB,\mathbb{R}^3)$ is the space of vector-valued functions in $\mathbb{R}^3$, i.e.\ $\bu_s$, defined on the manifold $\partial \mcB$ (the particle surface) whose euclidean norm is square integrable on $\partial \mcB$, i.e.\ $\int_{\partial \mcB}\|\bu_s\|_2^2\dd S < \infty$. We think this function space is a safe choice for $\bu_s$, but it might not be the most appropriate.},
where $\partial \mcB$ is the particle surface, which we will call \textit{``surface mobility operator"}, mapping the surface motion to the particle velocities. 
As shown by Swan \textit{et al.} \citep{Swan2011}, which generalizes the work of Stone and Samuel \citep{StoneSamuel1996},
$\wtil{\bN}$ can be obtained for a single particle using the Lorentz reciprocal theorem. \\

\itodo{
{\bf Modeling challenges:}\\
In order to model complex suspensions, 
it is crucial to develop methods that can capture short and long-ranged hydrodynamic interactions (HI) between tens or hundreds of thousands of particles.

However, complex particle shapes, non-trivial surface boundary conditions, the slow decay of HI in Stokes flow and their singular nature in the near-field make this task challenging. 

Traditional methods like the  Boundary Element Methods \citep{Pozrikidis1992} have been widely used to solve the many-body mobility problem in particle suspensions. However, they suffer from convergence issues at small interparticle distances,  require special, and complex, techniques to deal with the singularities distributed on the particle surfaces and are therefore limited to very few particles in dynamic simulations.



}

\subsection{Accounting for thermal fluctuations}
\label{sec:thermal_fluctu}
If the suspended particles are small enough, they will also move randomly due to thermal fluctuations in the surrounding fluid.  In order to satisfy the correct diffusion in the suspension, the statistics of the resulting particle fluctuating velocities, $\bU_{\text{th}}$,  must satisfy the fluctuation-dissipation theorem \cite{Kubo1966,Russel1981}
\eqn{
\avg{\bU_{\text{th}}} & =  \bzero,\\
\avg{\bU_{\text{th}}(t) \bU_{\text{th}}(t') ^T} &=  \bD \delta(t-t') = 2\kt \bN \delta(t-t'), 
 \label{eq:fluctu_dissip}
}
where the operator $\avg{\cdot}$ is the ensemble average,  $\bD$ the short-time diffusion tensor of the particles in the suspension, and $\kt$ the thermal energy.
Eq.\  \eqref{eq:fluctu_dissip} is a generalization of the famous Stokes-Einstein relation for a single particle to suspensions with multiple particles,
where the $6M\times6M$ short-time diffusion tensor of the particles $\bD$ is proportional to the mobility matrix.  
The fluctuations have the same origin as the viscous dissipation one must do work against when perturbing the system \cite{Groot1984},
and the fluctuation-dissipation theorem predicts that the response of a system in thermodynamic equilibrium to a small applied force is
the same as its response to a spontaneous fluctuation \cite{Guazzelli2011Book}.

In order to incorporate the effects of Brownian motion in the fluid,  a stochastic stress tensor $\bmZ$ \citep{Landau1959,Fabritiis2007} is added into the  right hand side of the Stokes equations
\eqn{
- \nabla p + \eta \nabla^2 {\bs{v}} &= \bna \cdot \bmZ, \label{eq:fluctu_Stokes1}\\
\nabla \cdot {\bs{v}} &= 0 \label{eq:fluctu_Stokes2},
}
The tensor $\bmZ$ encodes the stress generated by the random motion and collisions of the fluid molecules, which, at scales much larger than
the molecular scale, can be modeled as delta correlated in space and time
\eqn{
  \label{eq:avg_Z}
  \avg{\mc{Z}_{ij}(\bx, t)} &= 0,\\
  \label{eq:avg_ZZ}
  \avg{\mc{Z}_{ij}(\bx, t) \mc{Z}_{kl}(\bx', t^{'})} &= 2\eta \kt \pare{\delta_{ik}\delta_{jl} + \delta_{il}\delta_{jk}} \delta(\bx- \bx') \delta(t - t'),
}
The ensemble average of $\bmZ$ is zero because $\bmZ$ represents the random fluctuations around the mean deterministic flow \cite{Landau1959, Groot1984}.
The magnitude of the correlations, proportional to the thermal energy $\kt$, and their symmetries ensure that the fluctuation-dissipation theorem \eqref{eq:fluctu_dissip} is satisfied
by Eqs.\ \eqref{eq:fluctu_Stokes1}-\eqref{eq:avg_ZZ} \cite{Landau1959,Fabritiis2007}.



The mobility problem formed by \eqref{eq:fluctu_Stokes1}-\eqref{eq:fluctu_Stokes2} together with the boundary conditions \eqref{eq:no-slip-continuum} and force-torque balance \eqref{eq:balanceF_continuum}-\eqref{eq:balanceT_continuum} leads to the (It\^o) overdamped Langevin equations for the particle positions and orientations \citep{Hinch1975,noetinger1990fluctuating,roux1992brownian}
\begin{equation} 
     \bU = \bN \bF - \wtil{\bN}\bu_s  + \sqrt{2k_BT} \bN^{1/2}\bZ + k_BT\partial_{\bx}\cdot \bN,
    \label{eq:over_langevin}
\end{equation}
where two additional terms appear on the right hand side.

The first is the random particle velocity increments due thermal agitation given by $\bU_{\text{th}} = \sqrt{2k_BT} \bN^{1/2}\bZ$, with $\bN^{1/2}$ being the square root of the mobility matrix (i.e.\ $\bN^{1/2} \pare{\bN^{1/2}}^T = \bN$), and $\bZ$ a vector of independent white noise processes.
The noise dependence on the mobility matrix ensures that the fluctuation-dissipation theorem \eqref{eq:fluctu_dissip} is satisfied, a necessary condition for the Boltzmann distribution to be recovered at equilibrium \cite{Delong2014}. 
Note that while both $\bmZ$ and $\bZ$ present in Eqs.\ \eqref{eq:fluctu_Stokes1} and \eqref{eq:over_langevin} represent white noise terms they are not equivalent because the first is a continuous field defined on the fluid while the second is a discrete vector on the particles.

Along with the random velocities, the overdamped equations of motion require the inclusion of the thermal drift term, $k_BT\partial_{\bx}\cdot \bN$, which is the divergence of the mobility matrix with respect to the particle positions and orientations \citep{ermak_brownian_1978}. This ensures that the stochastic differential equation with an It\^o interpretation of the stochastic integral yields dynamics consistent with those of Smoluchowski's equation for the corresponding probability distribution.

Using uncontrolled approximations in the random particle velocities and drift terms results in incorrect equilibrium particle distributions and distorted dynamics which cannot always be predicted \textit{a priori}.
For example, simple approximations in the random particle velocities  can lead to errors of about $15\%$ in the diffusion coefficient of polymers and up to $50\%$ for the diffusion
of particles in dense suspensions \cite{Ando2012} (these errors were measured \emph{a posteriori} because  it was not possible to predict their magnitude beforehand). 
Neglecting the drift term leads to $\sim 40\%$ errors in the probability distribution function a single colloid diffusing inside a channel \citep{Delong2014,Delmotte2015,Sprinkle2019}
and result in incorrect dynamics of active particles near walls \cite{Leishangthem2024}.


\itodo{ 
{\bf Modeling challenges:}\\
Computing the square root of the mobility matrix $\bN^{1/2}$ with classical linear algebra, such as Cholesky decomposition, is costly as it scales badly with the particle number $O(M^3)$. Though several methods have been introduced to accelerate the matrix square root computation \citep{Fixman1986,Ando2012,Ando2013, Chow2014}, the inclusion of thermal fluctuations has often limited simulations to having very small particle numbers or ignoring completely hydrodynamic interactions between the particles. 

The additional drift term $k_BT\partial_{\bx}\cdot \bN$ requires computing the divergence of the mobility matrix with respect to \textit{all} the particle positions \textit{and} orientations, which incurs a high computational cost and cannot be done analytically for complex particle shapes.


In the past decade, significant advances have been made to compute these terms at a nearly linear computational cost ($O(M)$) in suspensions of spherical particles
\citep{Ando2012,Ando2013,Delong2014,Keaveny2014,Delmotte2015},
but their generalization to nonspherical shapes has been lacking, and traditional methods for more complex shapes, such as the Boundary Element Method 
\citep{Pozrikidis1992} or the Method of Regularized Stokesleta \citep{Cortez2001}, cannot handle thermal fluctuations.
}

\subsection{Constrained kinematics}
\label{sec:constraints}
%
%

In a variety of scenarios, the movement of certain bodies is either wholly or partially constrained, while others are free to move. This phenomenon can be observed in systems comprising freely-flowing particles and immobile obstacles, such as suspensions flowing through porous media or in microfluidic chips structured by pillar arrays.
 In other cases, which encompass a diverse array of microscopic systems, small objects are locked by geometric constraints or connected by rigid bonds with exceedingly short relaxation timescales. Stiff bonds can be found in polymer chains, freely-jointed molecules \citep{Sacanna2010,McMullen2018} or in molecules with bond angles. They can also be found in the hinges of multilink artificial microswimmers \citep{Dreyfus2005,Jang2015,Liao2019}, in the hook connecting the flagellum to the cell body of bacteria \citep{Berg1973,Trachtenberg1992}, or in the form of adhesive forces between sliding cells in diatom colonies \citep{Muller1782,Kapinga1992,Gordon2016}.  
The use of kinematic constraints allows for the elimination of these fast degrees of freedom, thereby enabling larger time steps in numerical simulations.

  In the majority of cases, the kinematic constraints mentioned above are holonomic, meaning that they depend only on the particle positions, orientations (and time $t$) and can be written in the form 
 \eqn{
 \bg(\bx,t) = \bzero
 }
 where we recall that $\bx = \{\bq,\btheta\}$ is a vector that collects all the particle positions and orientations.
 Typical holonomic, scalar or vectorial, constraints are for instance position constraints
 \eqn{
 \label{eq:g1} 
 \bg(\bx,t)  = \bq_p - \bq_0(t) = \bzero,}
 where $\bq_0(t)$  is the
  prescribed position  of a  body $p$;   
  prescribed distance $d(t)$  between two bodies  
 \eqn{\label{eq:g3} 
 g(\bx,t) = \fr{1}{2}\|\bq_p-\bq_q\|_2^2 - d(t)^2 = 0,}
  or ball-and-socket joints  between two bodies 
 \eqn{
  \label{eq:constraint}
  \bg(\bx) = \bq_p + \bR(\btheta_p) \bDl_{p}(t) - \bq_q - \bR(\btheta_q) \bDl_{q}(t) = \bs{0},
 }
 where $\bDl_{p}(t)$ is the vector from the body $p$ to the hinge of the joint in the body frame of reference. This vector is then rotated to the laboratory frame of reference by the
rotation matrix $\bR(\btheta_p)$. 
A constant vector $\bDl_{p} = \bs{c}$ represents a passive link, while $\bDl_{p}=\bDl_{p}(t)$ represents a moving link.

 Following the classical framework of Lagrangian mechanics, a Lagrange multiplier is associated to each kinematic constraint in the equations of motion. In the context of immersed bodies at low Reynolds number, the inclusion of Lagrange multipliers results in the introduction of constraint forces $\bF^C$ in the mobility problem \eqref{eq:U_NF}
 \eqn{
  \label{eq:U_NF_Fc}
  \bU = \bN (\bF+\bF^C) - \wtil{\bN}\bu_s.
}

In order to model suspensions with kinematic constraints, one has to solve the augmented mobility problem \eqref{eq:U_NF_Fc} together with the equations of motion  \eqref{eq:dx_dt}  subject to a diversity of kinematic constraints (e.g.\ \eqref{eq:g1}-\eqref{eq:constraint}) acting on a portion, or the totality, of the degrees of freedom in the system. Due to the nonlinearity of the kinematic constraints, this yields a nonlinear system of equations for the positions, orientations, as well as the constraint forces associated with each constraint.\\


\itodo{
{\bf Modeling challenges:}\\
 Nonlinearities pose a challenge to integrate the equations of motion \eqref{eq:dx_dt} for large suspensions \citep{Featherstone1987,Schoeller2021}.  
They require using nonlinear solvers, usually relying on Newton's method, where one needs to compute a large and complex inverse Jacobian matrix at each iteration. 
Until now, this challenging task has limited numerical studies to small numbers of particles \citep{Shum2010,Shum2017,Walker2019} or led to methods restricted to a specific type of constraint \citep{Schoeller2021}.
}

\subsection{Additional couplings: example with reactive  particles }
\label{sec:Laplace}

In addition to hydrodynamic interactions, suspended particles can interact through other interactions such as chemical, electrostatic or magnetic couplings. Just like the Stokes equations, the equations governing these fields are elliptic, which means that they can be solved with similar methods.  Bellow we outline the example of reactive phoretic particle.

The individual and collective dynamics of self-propelled reactive particles have attracted significant attention in recent decades  \citep{MoranPosner2017,Stark2018,Illien2017,Dominguez2022,Zottl2023}. 
Chemically-active phoretic colloids catalyze surface reactions to modify the concentration of chemical solutes surrounding them in order to self-propel, a phenomenon known as self-diffusiophoresis.  
In doing so, they  generate long-ranged hydrodynamic flows and chemical gradients that modify the trajectories of other particles. As a result, the dynamics of reactive suspensions is fundamentally governed by hydro-chemical interactions.

   Assuming that the magnitude of the phoretic flows is small and the solute diffusivity $D$ is large, the Péclet number characterizing the motion of the solute particles is small enough to neglect solute advection. Thus, in the fluid domain the concentration of solute  $c=c(\br)$ diffuses according to the Laplace equation 
\eqn{
  \label{eq:Laplace}
  D \bna^2 c = 0.
}

The immersed colloids activate chemical reactions on their surfaces to produce or consume solute. They are introduced as boundary conditions of the Laplace equation.
The boundary conditions are imposed on the concentration fluxes, i.e.\ Robin boundary conditions,
and to simplify the mathematical problem we only consider linear boundary conditions.
In particular, we consider sinks that consume solute at a rate $k$ proportional to the local concentration,
sources that produce solute at a constant flux  or a linear combination of both \citep{Golestanian2007,Lu1998, MichelinLauga2014}
\eqn{
  \label{eq:BC}
  D\frac{\partial c}{\partial \bn} \equiv D \bn \cdot \bna c = kc - \alpha \; \text{ on } \partial \mcB_p,
}
where $\bn=\bn(\br)$ is the surface normal on the colloids, $k=k(\br)>0$ the surface reaction rate and $\alpha=\alpha(\br)>0$ the surface production flux,
both of which can vary over the colloidal surface.
Since the reaction flux is proportional to the local concentration while the concentration diffuses with a finite diffusion coefficient $D$, this
boundary condition allows to model either diffusive-limited or reaction-limited systems.
The balance between diffusion and reaction is characterized by the
the Damkohler number, $\text{Da} = k R / D$, which represents the ratio between the diffusive time over one colloidal radius, $R^2/D$, and the reaction time scale $R/k$ \citep{Bhalla2013,MichelinLauga2014}.
Additional boundary conditions at the domain boundary, such as vanishing concentration at infinity, or zero flux at the domain walls, must be included to close the system.


Due to diffusiophoresis 
 the concentration gradients near the colloid surface induce  a slip velocity  \citep{Anderson1989,Marbach2019}
\eqn{
  \label{eq:slip}
  \bu_s = \mu \bna_{\parallel} c = \mu \pare{\bI - \bn\bn^T} \bna c \; \text{ on } \partial \mcB_p,
}
where the coefficient of proportionality $\mu=\mu(\br)$ is the surface mobility and $\bna_{\parallel}$ the surface gradient,
i.e.\ the gradient projected to the surface tangent plane. 

Modeling reactive suspensions requires solving first the Laplace problem \eqref{eq:Laplace}, with Robin boundary conditions on the particle surfaces \eqref{eq:BC}, in order to determine the concentration and then its gradients on the particle surfaces $\bna c$, which are used to compute the active slip \eqref{eq:slip} that appears in the boundary conditions of the (fluctuating) Stokes mobility problem \eqref{eq:over_langevin}.\\

\itodo{
{\bf Modeling challenges:}\\
The Laplace problem bear some similarities with the Stokes problem:  interactions are long-ranged and boundary conditions are non-trivial on the particles' surface.  Solving for the surface concentration gradients in large collections of (colloidal) reactive particles with arbitrary shapes and non-uniform catalytic surface properties is therefore challenging. Similarly to the Stokes problem, the preferred method for such systems is the Boundary Element, which does not handle thermal fluctuations and is restricted to very few particles in dynamic simulations.
}

\section{The rigid multiblob framework}
\label{sec:RMM}

As explained in Section \ref{sec: governing} the particle shape, hydrodynamic interactions, thermal fluctuations, constrained kinematics and additional non-hydrodynamic couplings pose multiple challenges to the modeling of complex particle suspensions.

In this section we present recent efforts to develop a unified, yet versatile, and scalable framework, based on the multiblob model, for modeling hydrodynamic and multiphysics interactions in large suspensions of arbitrarily shaped passive and active particles, taking into account thermal fluctuations and kinematic constraints.

\begin{figure}[t!]
  \centering
  \includegraphics[width=0.75 \textwidth]{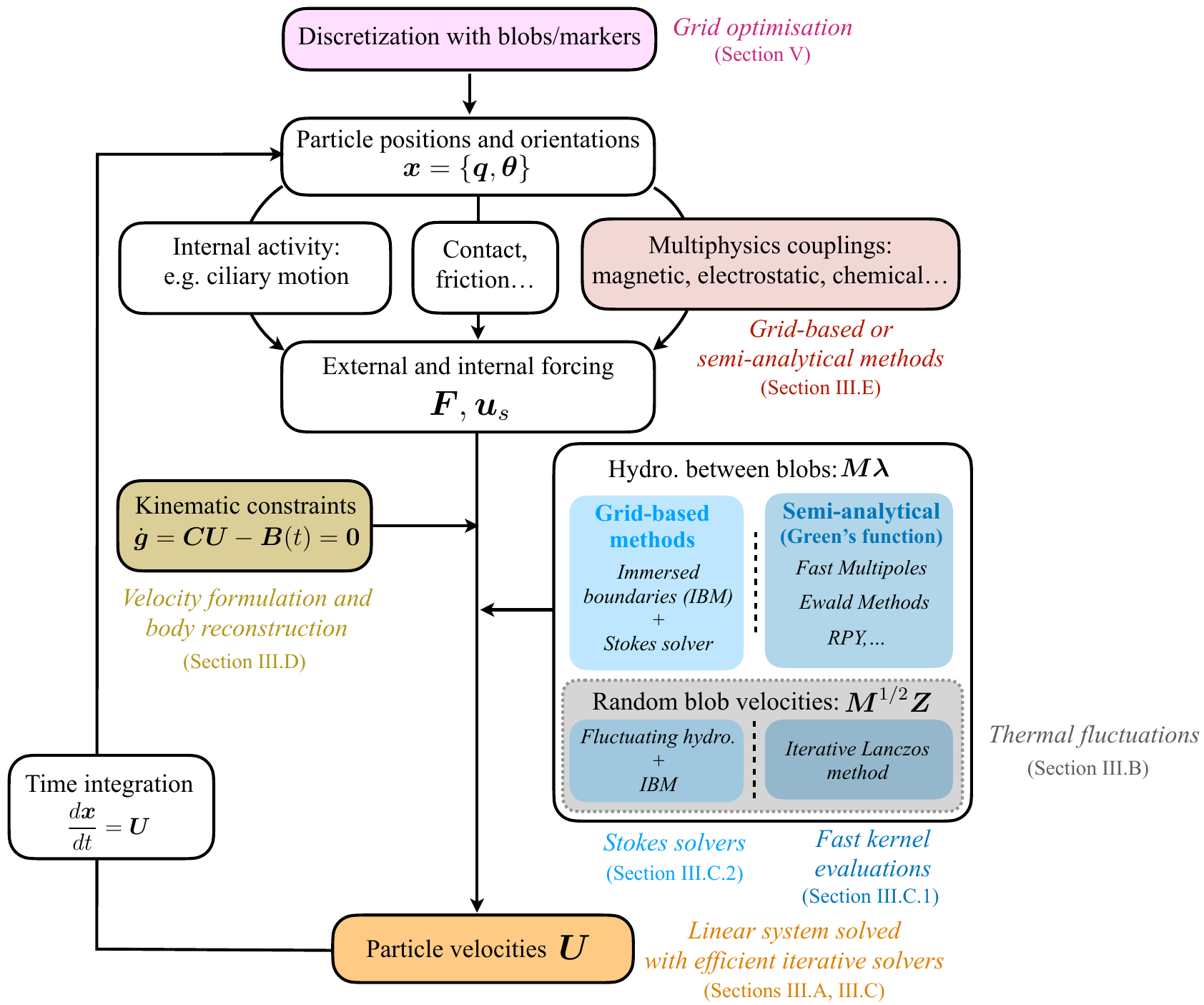}
  \caption{Diagram of the rigid multiblob (RMB)  framework. The colored boxes refer to a methodological challenge that is addressed in a subsection of the manuscript.
  }
  \label{fig:diagram_RMM}
\end{figure}

Figure \ref{fig:diagram_RMM} shows a diagram of the rigid multiblob framework for modeling complex suspensions. Each colored box represents a bottleneck in the method which has recently undergone significant development, and the text next to it refers to the corresponding section in this article.

\subsection{Rigid multiblob model for the Stokes problem}
\label{sec:RMM_eqs}

\begin{figure}[t!]
  \centering
  \includegraphics[width=0.9 \textwidth]{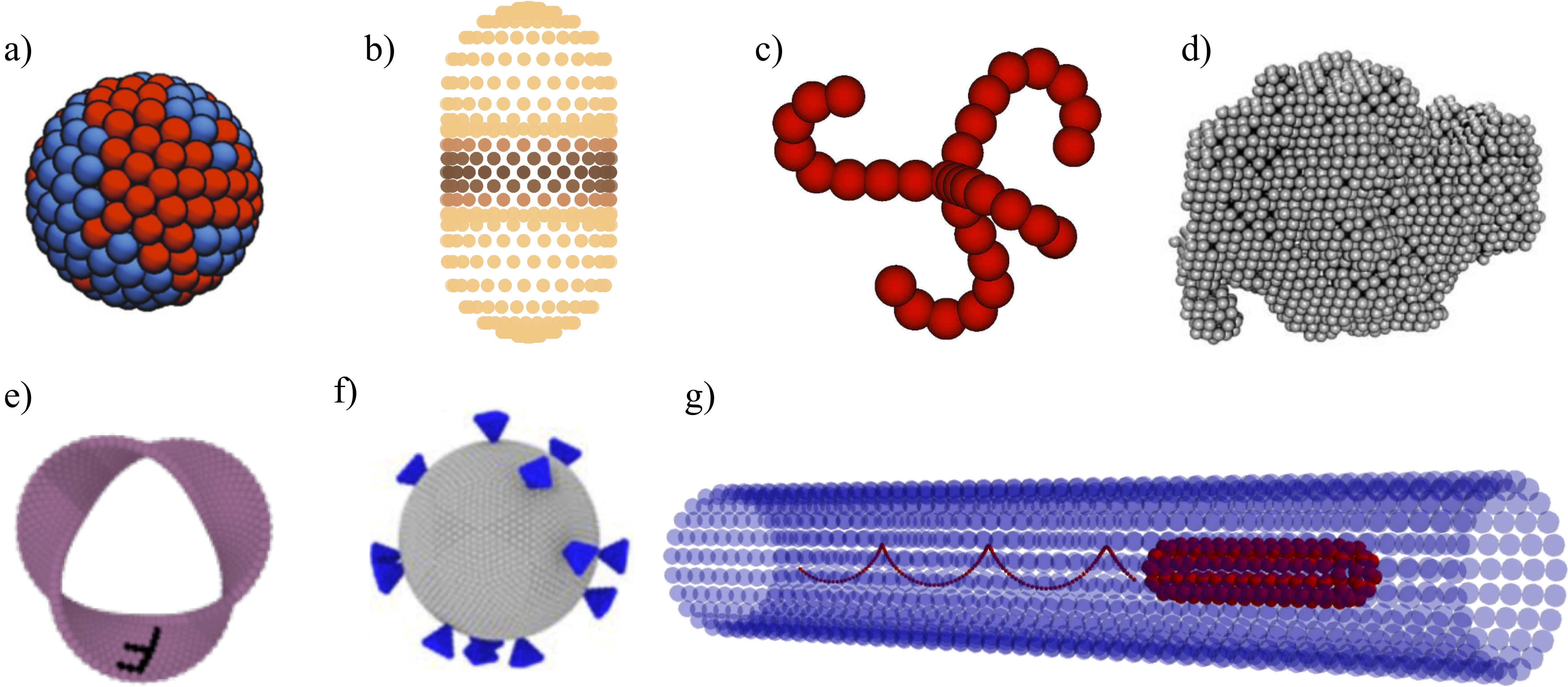}
  \caption{Examples of particles of various shapes discretized with the RMB model. (a) Nonuniform patchy sphere \cite{wang2019surface}. (b) Chemically active rod \cite{Delmotte2024}. (c) Branched particle \cite{Balboa2024}. (d) Globular protein called Lysozyme \citep{de2000calculation}. (e) M\"obius strip \citep{Moreno2024}. (f) HIV \citep{Moreno2022}. (e) Bacterium in a pipe \citep{Vizsnyiczai2020}.
  }
  \label{fig:ex_RMB}
\end{figure}

In the rigid multiblob (RMB) method rigid bodies are discretized using a collection of minimally resolved spherical nodes of hydrodynamic radius $a$, also called \emph{blobs}, which are constrained to move as a rigid body. 
Figure \ref{fig:ex_RMB} shows examples of various geometries discretized with the RMB.

Each blob is subject to a force, $\blambda_i$, that enforces the rigid motion of the body. 
Thus, the integrals in the balance of force and torque \eqref{eq:balanceF_continuum}-\eqref{eq:balanceT_continuum} become sums over the blobs
\eqn{
\label{eq:balanceF}
\sum_{i\in \mcB_p} \blambda_i   &= \bbf_p, \\
\label{eq:balanceT}
\sum_{i \in \mcB_p} (\br_i - \bq_p) \times \blambda_i  &= \btau_p,
}
where $\br_i$ is the position of blob $i$ belonging to body $p$ ($\mcB_p$). Note that here $\blambda_i$  now represents  finite forces and not density forces as in the continuum formulation since it includes  the quadrature weights of the blobs. 
The boundary condition on the particle surfaces \eqref{eq:no-slip-continuum}, as in collocation methods \citep{Pozrikidis1992}, is evaluated at each blob
\eqn{
\label{eq:no-slip}
\bv(\br_i) =  \sum_j \bM_{ij} \blambda_j &= \bu_p + \bomega_p \times (\br_i - \bq_p)  + \bu_s(\br_i) \;\; \mbox{for all } i \in \mcB_p, 
}
where $\bv(\br_i)$ is the fluid velocity at the position of blob $i$, $\bM$ is the $3N_b\times 3N_b$ mobility matrix that mediates the hydrodynamic interactions between blobs, and $N_b$ is the total number of blobs in the system.
The term $\bM_{ij}$ couples the force acting on the blob $j$ to the velocity of the blob $i$. 
Depending on the geometry of the domain under consideration, this matrix can be written explicitly if the Green's function of the problem is known, or its action can be computed with matrix-free methods relying on grid-based Stokes solvers, such as Immersed Boundary Methods \citep{Peskin2002} or the  Force Coupling Method \citep{Maxey2001} (see Section \ref{sec:FCM}).

A well-known example of regularized Green's function is the Rotne-Prager-Yamakawa (RPY) approximation \citep{Rotne1969, Wajnryb2013}, where the regularization is done by a double convolution of the Stokes Green's function, $\bmG(\br, \br')$, with Dirac delta functions defined on the surface of a sphere of radius $a$:
\eqn{
  \label{eq:RPY}
  \bM_{ij} = \bM(\br_i, \br_j)  &=  \fr{1}{(4\pi a^2)^2}  \int \delta(|\br' - \br_i|-a) \bmG(\br', \br'') \delta(|\br'' - \br_j|-a) \dd^3 r'' \dd^3 r'.
}
Wajnryb et al.\ showed that, for non-overlapping blobs, an equivalent definition of the RPY mobility is \citep{Wajnryb2013}
\eqn{
  \label{eq:RPY_v2}
  \bM_{ij} = \bM(\br_i, \br_j) = \corchete{\bI + \fr{a^2}{6} \bna^2_{\br'}} \corchete{\bI + \fr{a^2}{6} \bna^2_{\br''}} \bmG(\br', \br'') \vert_{\br''=\br_j}^{\br'=\br_i},
}
which makes evident the role of the blob radius as a regularization parameter.

The well-known RPY mobility matrix was developed to provide a symmetric positive definite, pairwise approximation.
The use of a symmetric positive definite matrix $\bM$ is one of the main advantages of the RMB as it allows to 
incorporate Brownian fluctuations efficiently as  shown in Sec.\ \ref{sec:therm_fluctu}.
In contrast, the Regularized Stokeslets method uses a mobility $\bM$ that is not positive definite by design \cite{Cortez2005} and,
current Boundary Element Methods use singular and near singular quadrature rules to compute the action of the Stokes Green's function, $\bmG(\br, \br')$
that generate non-symmetric mobility matrices $\bM$ and thus, cannot incorporate Brownian fluctuations efficiently
(see Ref.\ \cite{bao2018fluctuating} for an exception in 2D). As explained in Sec.\ \ref{sec:FCM}, the mobility matrix of grid-based- methods is symmetric positive definite by design and has the same structure \eqref{eq:RPY_v2} as the RPY mobility to order $\mc{O}\pare{a^4}$.

Since $\bM$ is positive definite it is not necessary to employ special quadrature rules to evaluate near or self interactions.
However, the formula \eqref{eq:RPY} introduces a regularization error proportional to the blob radius \cite{Delmotte2024}.
As a typical discretization of a rigid body surface employs $N_b \sim 1/a^2$ blobs, the convergence rate scales with the number of blobs as $\sim N_b^{-1/2}$.
Despite this slow convergence it is possible to use optimized grids to obtain results accurate up to a few percent as far
as blobs of different bodies do not overlap \cite{Broms2023, Delmotte2024}.
We illustrate the accuracy of the RMB method in a couple of examples in Secs.\ \ref{sec:articulated}-\ref{sec:multi-physics}
and discuss optimized grids in Sec.\ \ref{sec:perspectives}.

The original RPY mobility computes hydrodynamic interactions between spherical particles of equal radii in an unbounded domain \citep{Wajnryb2013}.
Extensions of the RPY matrix for particles of different radii \citep{Zuk2014}, in a background shear flow \citep{Wajnryb2013}, above a fluid-fluid interface \citep{Delmotte2023} or a no-slip boundary \citep{Swan2007} with arbitrary periodicities \citep{Yan2018,Yan2021}, or in a spherical cavity \citep{Aponte2016} are available in the literature. 
Section \ref{sec:RPY_Lanczos} enumerates fast methods to compute the action of the RPY tensor.



The whole linear system to solve the  mobility problem is found by combining \eqref{eq:balanceF}-\eqref{eq:no-slip},
\eqn{
\label{eq:linear_system}
\left[\begin{array}{cc}
\bM & -\bK  \\
-\bK^T & \bzero  
\end{array} \right]
\left[\begin{array}{c}
\blambda \\
\bU 
\end{array} \right] =
\left[\begin{array}{c}
\bu_s \\
-\bF 
\end{array} \right].
}
In the linear system the matrix $\bK$ is a discretization of the operator $\bs{\mc{K}}$ introduced in Section \ref{sec:governing_HI}: it transforms the rigid body velocities to blob velocities.
The solution of \eqref{eq:linear_system} is 
\eqn{
\label{eq:mob_sol_multiblob}
\bU = 
\bN\bF - \bN\bK^T\bM^{-1}\bu_s
}
where $\bN = \left(\bK^T\bM^{-1}\bK\right)^{-1}$ and $\wtil{\bN}=\bN\bK^T\bM^{-1}$ are the $6M\times 6M$ body mobility matrix and $6M\times 3N_b$ surface mobility operator given by the rigid multiblob model. \\

\itodo{
{\bf Key contributions:}\\
The RMB allows to simulate suspensions of bodies of arbitrary shape,  with surface velocities, without dealing with singularities, in a robust, simple and versatile framework.
The RMB naturally interfaces with \textit{any} fast method, either grid-based or relying on Green's functions, for computing hydrodynamic interactions between blobs (see Section \ref{sec:precond_fast_HI}).
The linear system \eqref{eq:linear_system} can be solved efficiently with preconditioned iterative methods (see Section \ref{sec:precond_fast_HI}).
}

\subsection{Including thermal fluctuations}
\label{sec:therm_fluctu}

As explained  in Section \ref{sec:thermal_fluctu}, two new terms appear in the equations of motion when thermal fluctuations are included: the stochastic particle velocities $\bU_{\tex{th}} = \sqrt{2k_BT}\bN^{1/2}\bZ$ and the drift term $k_BT\partial_{\bx}\cdot \bN$.
Both of these terms are challenging to account for with non-spherical particles and are computationally expensive. 
Below we present the formalisms recently developed to  handle these terms correctly and efficiently.

\subsubsection{Stochastic particle velocities}
The computation of the stochastic particle velocities $\bU_{\tex{th}}=\sqrt{2k_BT}\bN^{1/2}\bZ$ can be readily incorporated into the system by including a random velocity at the blob level, $\bu_{\tex{th}}$, in the boundary conditions \eqref{eq:no-slip} \citep{Sprinkle2017,Westwood2022}.  Specifically, we consider the fluctuating mobility problem 
\eqn{
  \label{eq:linear_fluctu}
  \begin{bmatrix}
    \bM & -\bK \\
    -\bK^T & \bzero
  \end{bmatrix}
  \begin{bmatrix}
    \blambda \\
    \bU
  \end{bmatrix}
  =
  \begin{bmatrix}
     \bu_s - \bu_{\tex{th}}\\
    -\bF
  \end{bmatrix},
}
with the random blob velocity,
\eqn{
  \label{eq:u_th} 
  \bu_{\tex{th}} = \sqrt{2 \kt} \bM^{1/2} \bZ,
}
where $\bM^{1/2}$ is the \emph{square root} of the blob mobility matrix, i.e.\ $\bM^{1/2}\pare{\bM^{1/2}}^T=\bM$, and $\bZ$ is a $3N_b\times 1$ white noise vector acting on the blobs, with covariance $\avg{\bZ(t) \bZ^T(t')} = \bI \delta(t - t')$.
This form of the noise term is equivalent to including a stochastic stress directly into the Stokes equation as in  \eqref{eq:fluctu_Stokes1}.
Since with the RPY approximation and grid-based approaches the mobility $\bM$ is positive definite \cite{Delong2014,Delmotte2015}, the stochastic velocity \eqref{eq:u_th} is well defined. 
The covariance of the resulting rigid body velocities generated by the stochastic blob velocities, setting $\bF=\bzero$ and $\bu_s=\bzero$, is 
\eqn{
  \avg{\bU \bU^T} &= \avg{\bN\bK^T\bM^{-1}\bu_{\tex{th}}\bu_{\tex{th}}^T\bM^{-1}\bK\bN} \nonumber\\
      &= (\bN \bK^T \bM^{-1})  \sqrt{2\kt} \bM^{1/2} \avg{ \bZ \bZ^T } \bM^{1/2}  \sqrt{2\kt} (\bN \bK^T \bM^{-1})^T  \nonumber  \\
    &= 2\kt \bN \bK^T \bM^{-1}  \bM  \bM^{-1} \bK \bN  \,\delta(t - t') \nonumber  \\
    &= 2\kt \bN (\bK \bM^{-1} \bK^T) \bN  \,\delta(t - t')  \nonumber  \\
  &= 2\kt \bN \,\delta(t - t') = \avg{\bU_{\text{th}} \bU_{\text{th}} ^T}.
  \label{eq:fluctu-diss}
}
Eq.\ \eqref{eq:fluctu-diss} shows that the thermal noise is balanced by the viscous dissipation as required by the fluctuation-dissipation theorem \eqref{eq:fluctu_dissip}.
Thus the stochastic velocity imposed at the blob level \eqref{eq:u_th} generates Brownian body velocities consistent with the Stokes equation. 
Equivalently, the square root of the body mobility matrix  $\bN^{1/2} = \bN\bK^T\bM^{-1/2}$ satisfies the fluctuation-dissipation theorem with $\bN^{1/2}(\bN^{1/2})^T = \bN$.

\subsubsection{Drift term}
The second aspect of the computation is to account for Brownian drift when advancing the particle positions. \\
It has been shown \citep{Delong2014, Sprinkle2017,Sprinkle2019} that the drift can be incorporated using random finite differences (RFD), which a randomized version of the standard finite differences: given a Gaussian random vector, $\bW$, such that $\langle \bW \bW^T \rangle =  \bI$, the divergence of the mobility matrix can be approximated as follows
\eqn{
  k_BT\partial_{\bx}\cdot \bN = k_BT \left\langle \fr{\bN\pare{\bx + \delta\bW/2}-\bN\pare{\bx - \delta\bW/2}}{\delta}\cdot  \bW\right\rangle + O(\delta^2),
}
where $\delta$ a \textit{small} number.
This involves applying the mobility matrix $\bN$, and thus solving two mobility problems \eqref{eq:linear_fluctu},
evaluated at randomly displaced positions/orientations $\pm\delta\bW / 2$ to a random force/torque vector $k_BT\bW$.
Since the parameter $\delta$ is small the solution of one mobility problem can be used as initial guess to the second mobility problem reducing the computational cost \cite{Fiore2019}.

In addition to RFD, a new midpoint scheme, called the ``generalized Drifter-Corrector'' (gDC), was recently developed \cite{Westwood2022}.  This scheme, which is ideally suited for grid-based solvers, \textit{does not require} additional mobility solves compared to the deterministic case and exhibit the same accuracy as RFD.   \\

\itodo{
{\bf Key contributions:}\\
Introducing the noise at the blob level allows to generate the correct Brownian velocity increments at the level of the bodies with arbitrary shapes.
The random blob velocities can be computed  with \textit{any} fast method, either grid-based or  Green's function-based (see Section \ref{sec:precond_fast_HI}).
Thanks to recent mathematical advances, including the drift term only requires one to two solves of the mobility problem \eqref{eq:linear_fluctu} per time step. The resulting  computational cost is negligible compared to traditional schemes, such as Fixman's method \cite{Fixman1978,Fixman1986}. 
}

\subsection{How to scale big and compute fast?}
\label{sec:precond_fast_HI}

The linear system \eqref{eq:linear_fluctu} needs to be solved efficiently to carry out dynamic simulations with large numbers of particles. To do so, two ingredients are crucial:
\begin{enumerate}
 \item \textbf{An efficient iterative solver.} The size and condition number of the linear system \eqref{eq:linear_fluctu} grows with the number of blobs $N_b$, it is therefore  essential to use an iterative solver with a convergence rate that depends weakly on the number of particles. For that purpose block diagonal preconditioners applied to the iterative GMRES method have been developed \cite{Usabiaga2016,Fiore2019,Usabiaga2022}. The resulting solvers exhibit a convergence within a few iterations that is \textit{independent} of the particle number $M$.
 
 \item \textbf{Fast hydrodynamic interactions between blobs.}  Hydrodynamic interactions between bodies are obtained directly  from the matrix $\bM$.  The action of $\bM$ on a vector is required to compute the deterministic velocities $\bU = \bN\bF - \wtil{\bN}\bu_s$, while the product $\bM^{1/2} \bZ$ is necessary to obtain the Brownian velocities $\bU_{\tex{th}}$. It is therefore essential to use and develop efficient methods to compute the action of $\bM$ and its square root. In the following, we describe the approaches that have been developed for the fast evaluation of $\bM$ using Green's function and grid-based methods.
\end{enumerate}

\subsubsection{Fast hydrodynamics with Green's functions}
\label{sec:RPY_Lanczos}
Using direct linear algebra, the computational cost of a mobility-vector products $\bM\blambda$ scales quadratically with the number of blobs ($O(N_b^2$)) with Green's function based methods. In order to reduce this cost,  matrix-vector products can be parallelized, either on CPU's or on the shared memory of  GPU's, which yields a linear scaling up to $N_b \sim 10^4-10^5$ \cite{Usabiaga2014,liu2014large,Sprinkle2017,Singh2020,Townsend2024,pelaez2025universally, torre2025python}.  More sophisticated methods are also used, such as fast multipole methods (FMM) \citep{Liang2013}, to achieve a linear scaling.  For periodic domains, the positively split Ewald method \citep{Fiore2017,Fiore2018}, the fast FCM \cite{su2024accelerating} and periodic FMM \citep{Yan2018,Yan2021} are the most efficient and scalable methods for  computing the action of $\bM$ on a vector.\\
 To compute the action of $\bM^{1/2}$ on the random vector $\bZ$, iterative methods such as the Lanczos algorithm are best suited \citep{Ando2012,Chow2014}. This method iteratively approximates $\bM^{1/2} \bZ$  with a set of basis vectors of a vectorial space $\mathbb{K}$ that are linear combinations of the powers of the mobility matrix times the random vector: $\mathbb{K}=\mbox{span}\left\{\bZ,\bM\bZ,\bM^2\bZ,...\right\}$.
The cost of the method therefore depends on the number of basis vectors, and thus mobility-vector products, required to reach a given tolerance $\epsilon$. Using a block-diagonal preconditioner, the convergence rate is independent of the number of blobs and a small number of iterations (typically 3-5) is required to reach a reasonable tolerance.

\subsubsection{Fast hydrodynamics with grid-based methods}
\label{sec:FCM}
When there is no explicit expression for the Green's function of the system, 
 matrix-free methods can be used.  They  rely on grid-based Stokes solvers  with fluctuating hydrodynamics combined with a large family of Immersed-Boundary (IBM)  methods  \citep{Atzberger2007,Atzberger2011, Delong2014, Keaveny2014, Plunkett2014, Sprinkle2019,Hashemi2023,pelaez2025universally}. \\
With fluctuating IBM methods, the coupling between the blobs and the fluid is achieved through a forcing term $\bbf$ added to the fluctuating Stokes equations \eqref{eq:fluctu_Stokes1},
\begin{align}
 \bna p - \eta \bna^2\bv &= \bna\cdot\bmZ + \bbf, \label{eq:fluc_stokes1}\\
 \bna\cdot\bv &= 0.
 \label{eq:fluc_stokes2}
\end{align}

The second term on the RHS of \eqref{eq:fluc_stokes1} is the blob forcing transferred to the fluid with a spreading operator $\mc{S}$,
\begin{align}
 \bbf(\br) = \mc{S}[\blambda](\br) = \sum_{i=1}^{N_b}\blambda_i\Delta_i\left(\br\right),
 \label{eq:spreading}
\end{align}
where the finite size of the blobs is accounted for with a spreading envelope $\Delta_i\left(\br\right)$. In the context of IBM, $\Delta_i\left(\br\right)$ is a piece-wise regular compact kernel  \citep{Peskin2002}. Other IBM-like methods, such as the Force Coupling Method (FCM), use Gaussian kernels \citep{Maxey2001}. 

The fluid velocity, obtained after solving \eqref{eq:fluc_stokes1} and  \eqref{eq:fluc_stokes2}, is the sum of a deterministic part $\bv^D$, due to the forcing $\blambda$, and a fluctuating term $\bv_{\tex{th}}$, stemming from the fluctuating stress.  We may express the total fluid velocity as
\begin{align}
 \bv &= \mc{L}^{-1}\left(\mc{S}\blambda + \mc{D}\bmZ\right)\\
 &= \bv^D + \bv_{\tex{th}},
\end{align}
 where $\mc{S}$ is the spreading operator in \eqref{eq:spreading}, $\mc{L}^{-1}$ is the inverse Stokes operator (i.e.\ the fluid solver), and $\mc{D}$ the divergence operator applied to the fluctuating stress \citep{Delong2014}.

The velocities of the blobs are then obtained from the fluid velocity using an averaging operator, $\mc{J}$, such that
\begin{align}
 \bu_i = (\mc{J}[\bv])_i = \int \bv\Delta_i(\br) \dd^3\br = \bu_i^D + \bu_{\tex{th},i}, \quad i=1,..,N_b,
 \label{eq:averaging}
\end{align}
where $\mc{J} = \mc{S}^T$ is adjoint to the spreading operator.

We see then that the IBM mobility matrix can be written as the composition of three linear  operators
\begin{align}
\bM_{\text{IBM}} = \mc{J}\mc{L}^{-1}\mc{S}.  
\end{align}
The IBM mobility differs from the RPY mobility defined in Eq.\ \eqref{eq:RPY}.
However, one can perform a second order Taylor expansion of the flow field inside the kernels to show that $\bM_{\text{IBM}}$ obeys \eqref{eq:RPY_v2} to order $\mc{O}\pare{a^4}$.
Thus, different kernels  define different regularizations  that are all essentially equivalent to the RPY mobility \cite{Delong2014,Kallemov2016}. 

Additionally, as demonstrated in \citep{Delong2014, Keaveny2014, Delmotte2015} the velocity $\bu_{\tex{th}}$ satisfies the fluctuation–dissipation theorem with the covariance given by the IBM  approximation of the mobility matrix, 
\begin{align}
 \left\langle \bu_{\tex{th}}(t)\bu_{\tex{th}}(t') \right\rangle= 2k_BT\bM_{\text{IBM}}\delta(t-t'),
 \label{eq:fluctu_dissip_blob}
\end{align}
where $\bu_{\tex{th}} = \sqrt{2k_BT}\bM^{1/2}_{\text{IBM}}\bZ(t)$, with $\bM^{1/2}_{\text{IBM}} =\mc{J}\mc{L}^{-1}\mc{D}$.
To fulfill \eqref{eq:fluctu_dissip_blob}, it is crucial that the averaging and spreading operators are adjoint ($\mc{J} = \mc{S}^T$)
\footnote{It is also necessary that the discretization of the Stokes equations preserves some of the original symmetries,
specifically $\mc{L}^{-1} \mc{D} \mc{D}^T \mc{L}^{-1} = \mc{L}^{-1}$, which is true in the continuum setting but it is not respected by all discretizations \cite{Delong2014}.}, which is not the case for the method of Regularized Stokeslets \cite{Cortez2005}.

Thus, we see that fluctuating IBM methods simultaneously compute the actions of $\bM$ and $\bM^{1/2}$ on $\blambda$ and $\bZ$, respectively, by considering solutions to the forced fluctuating Stokes equations.

The fluctuating IBM can be implemented with any kind of Stokes solver.  This involves first evaluating the forcing, $\bbf(\br)$, and the fluctuating stress, $\bmZ$, on the grid.  Next, the Stokes equations are solved with a Stokes solver (e.g.\ using Finite Volumes, Finite Elements, Finite Differences or a Spectral method).  Finally, a quadrature rule is used to numerically integrate \eqref{eq:averaging} to obtain the translational velocity for each of the blob making up the rigid bodies. Since many  Stokes solvers can be well-parallelized on processor cores or GPUs, the computational cost scales linearly with the  number of blobs $O(N_b)$ \citep{Delmotte2015c,Hashemi2023,pelaez2025universally}.\\

\itodo{
{\bf Key contributions:}\\
For both Green's function- and grid-based approaches, there are  fast, parallel, methods for computing hydrodynamic interactions between blobs and their random noise,  with linear scaling $O(N_b)$.
Thanks to efficient preconditioned iterative solvers, the rigid multiblob linear system \eqref{eq:linear_fluctu} can be solved with just a few iterations, i.e.\ mobility-vector products $\bM\blambda$, independently of the number of blobs in the system. Therefore the linear scaling of the computational cost is preserved at the level of the rigid bodies $O(M)$, which allows for simulations of complex suspensions at scales that had not been reached before.
}

\subsection{Handling kinematic constraints in large suspensions}
\label{sec:articulated}

As explained in Section \ref{sec:constraints}, handling nonlinear holonomic constraints in large suspensions of bodies with arbitrary shape is challenging.
However, when the constraints are holonomic their time derivatives are \textit{linear} in the velocities of the rigid bodies.
Therefore the time derivative of these constraints can be written compactly as a constraint equation on the particle \textit{velocities} 
\eqn{
  \label{eq:dg_compact}
  \dot{\bg} = \bC \bU  - \bB(t) = \bs{0},
}
where the matrix $\bC = (\partial \dot{\bg} / \partial \bU)$ is the constraints' Jacobian. 
For instance, the matrix $\bC$ and vector $\bB$ of the time derivative of the ball-and-socket constraint \eqref{eq:constraint} are given by
\eqn{
  \bC &= \left[\bI_{3\times 3}\;\; -\Delta\bl_{p}^\times\;\;-\bI_{3\times 3}\;\; \Delta \bl_{q}^\times\right], \nonumber \\
  \bB(t) &= \bR(\btheta_q) \dot{\Delta \bs{l}}_{q}(t) - \bR(\btheta_p) \dot{\Delta \bs{l}}_{p}(t), \nonumber
}
where $\Delta \bl_{q}^{\times} \by = \Delta \bl_{q} \times \by$ for any vector $\by$ and $\dot{\Delta \bs{l}}_{p}(t)=\frac{\dd \Delta \bs{l}_{p}}{ \dd t}$.

\begin{figure}[t!]
  \centering
  \includegraphics[width=0.85 \textwidth]{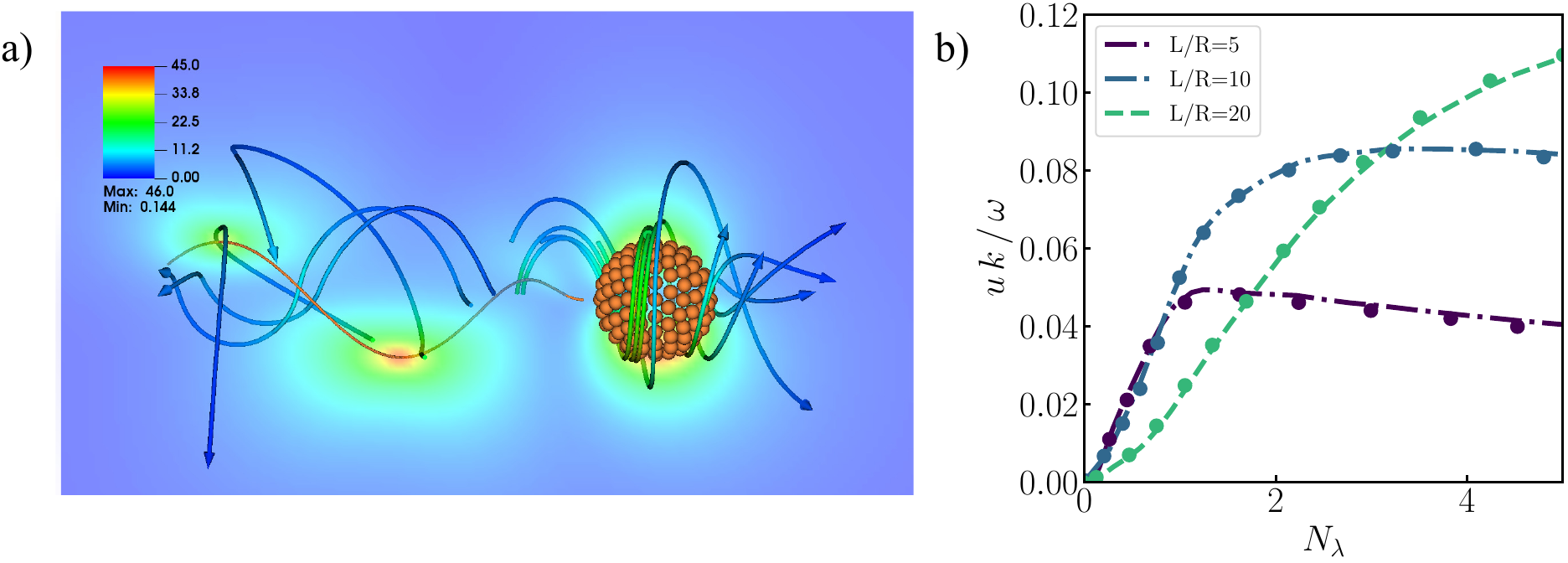}
  \caption{Illustration and validation of the RMB method with kinematic constraints. (a) Streamlines and fluid velocity magnitude around a swimming bacterium in free space discretized with 200 blobs. (b)  Swimming speed of the bacterium normalized by the helical wave velocity $\omega /k$, where $k$ is the wavenumber of the helical wave and $\omega$ the rotation frequency, as a function of the number of wavelengths $N_{\lambda}$ along the helical tail for different ratios of flagellar length $L$ to head radius $R$. Dashed line: RMB  method with kinematic constraints. Symbols: simulations from Higdon \citep{Higdon1979}. Figure adapted from \cite{Usabiaga2022}.
  }
  \label{fig:bacteria_fields}
\end{figure}

Each constraint exerts a force and torque on the rigid bodies which in the case of passive links, $\bC \bU=\bB(t)=0$, generate no work \citep{Goldstein2002}.
This condition is enough to determine the structure of the constraint generalized forces,
\eqn{
\bF^C = \bC^T \bphi
\label{eq:Fc}
}
where $\bphi$ is a Lagrange multiplier homogeneous to a force \citep{Delmotte2015b}.

Adding the constraint forces, $\bF^C = \bC^T \bphi$, in the mobility relation \eqref{eq:mob_sol_multiblob}, and inserting the velocity constraint \eqref{eq:dg_compact} yields the linear system for the Lagrange multipliers $\bphi$ and particle velocities $\bU$ \cite{Usabiaga2022}
\eqn{
  \label{eq:linear_system_general_const}
\left[\begin{array}{cc}
-\bN \bC^T& \bI  \\
 \bzero & \bC 
\end{array} \right]
\left[\begin{array}{c}
\bphi \\
\bU 
\end{array} \right] =
\left[\begin{array}{c}
\bN \bF - \wtil{\bN}\bu_s\\
\bB
\end{array} \right],
}
where $\bI$ is a $6M \times 6M$ identity matrix.
Any  method to solve the Stokes problem, and therefore to apply the mobility matrix $\bN$, can be used with \eqref{eq:linear_system_general_const}. A similar framework to handle holonomic constraints in particle suspensions has been developed by \cite{funkenbusch2024approaches} but with a different formulation of the linear system.


Solving \eqref{eq:linear_system_general_const} directly would require, in general, using nested linear solvers;
an inner solver to compute the action of $\bN$ and an outer solver to solve \eqref{eq:linear_system_general_const}.
Such approach would be computationally expensive.
However, with the rigid multiblob model presented above, the whole linear system to solve the  constrained mobility problem is found by combining \eqref{eq:balanceF}-\eqref{eq:no-slip}, \eqref{eq:dg_compact} and \eqref{eq:Fc} \cite{Usabiaga2022}
\eqn{
\label{eq:linear_system_const}
\left[\begin{array}{ccc}
\bM & -\bK & \bzero \\
-\bK^T & \bzero & \bC^T \\
\bzero & \bC & \bzero 
\end{array} \right]
\left[\begin{array}{c}
\blambda \\
\bU \\
\bphi
\end{array} \right] =
\left[\begin{array}{c}
\bu_s \\
-\bF \\
\bB
\end{array} \right].
}

This linear system has the same structure as the unconstrained problem \eqref{eq:linear_system}, which allows to use the same preconditioned iterative solvers  (see Section \ref{sec:precond_fast_HI}). 
While resistance problems are known to scale poorly with the particle number, the iterative solver for the \textit{mixed mobility-resistance problem} \eqref{eq:linear_system_const} is not sensitive to the system size, therefore allowing to study large suspensions of constrained bodies  with quasilinear computational cost.

Additionally, constraint violations, e.g.\ due to discrete time-integration errors, are prevented  with a body reconstruction that uses the constraints and a correction of the particles' positions and orientations at the end of each time-step. The correction procedure, based on a nonlinear minimization algorithm, has negligible computational cost and preserves the accuracy of the time-integration scheme \cite{Usabiaga2022}.

As a validation example, Figure \ref{fig:bacteria_fields}a  shows the flow field around a bacterium assembled with a spherical head and a helicoidal tail, both discretized with the multiblob method, that are attached and constrained to rotate in opposite directions along the same rotation axis.  The resulting swimming speed shown in Figure \ref{fig:bacteria_fields}b compares very well with other simulation results from the literature \citep{Higdon1979}.\\

\itodo{
{\bf Key contributions:}\\
A velocity formulation of the kinematic constraints between bodies enables to write the constrained problem  as \textit{a single linear system}.
This linear system is solved with a preconditioned iterative solver that couples effectively with any numerical method to compute hydrodynamic interactions between rigid bodies.
The solver convergence is independent of the system size and constraint type, therefore  allowing to simulate large suspensions in a scalable fashion.
The resulting tool can handle different populations of constrained bodies simultaneously where both passive and time-dependent constraints are directly read from a simple  input file.
Thanks to its simplicity and flexibility, the user can readily use it to study physical and biological systems involving large collections of constrained bodies.
}

\subsection{The multi-physics multiblob method}
\label{sec:multi-physics}

\begin{figure}[t]
  \centering
  \includegraphics[width=0.7 \textwidth]{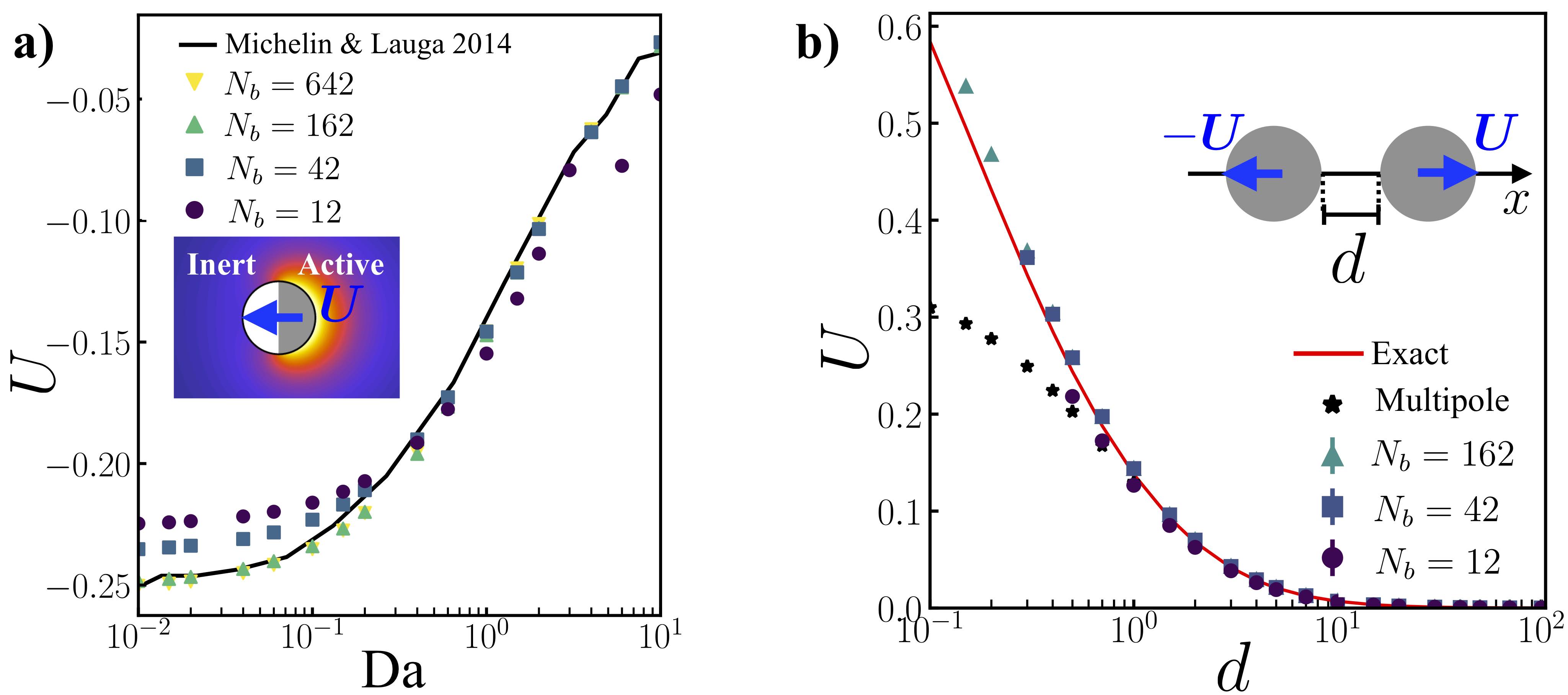}
  \caption{ Illustration and validation of the RMB method for reactive particles. a) Swimming speed for a Janus sphere as of function of the Damkohler number.
    Results for resolutions with different number of blobs, $N_b$, and comparison with the results of Michelin and Lauga \citep{MichelinLauga2014}.
    The results with 42 blobs agree well for high Da numbers and shows an error of about $5\%$ for $\text{Da} \ll 1$;
    for finer resolutions the agreement is good for all Da numbers.  Inset shows a the concentration field around a Janus particle (adapted from \citep{MichelinLauga2014}). b) Velocities along the $x$-axis for two spheres with uniform surface activity as a function of the gap $d$ between the spheres. 
    The motion is symmetric, thus only  the velocity magnitude is shown.
    The results for discretizations with $N_b=12,\,42 \text{ and } 162$ are compared with multipolar expansions \cite{Rojas2021} and the exact solution computed  in bispherical coordinates. Inset: sketch of the configuration. Figure adapted from \cite{Delmotte2024}.
  }
  \label{fig:valid_phoretic}
\end{figure}

Similarly to the Stokes equations, the Laplace equation \eqref{eq:Laplace} is an elliptic partial differential equations that can be written with an integral formulation at the particles' surfaces  \citep{Montenegro2015}  and  discretized with the multiblob grid.
The method has recently been extended to account for hydrochemical and phoretic interactions in suspensions of reactive particles with arbitrary shapes \cite{Delmotte2024}.

Just like the hydrodynamic problem, the multiblob method for the Laplace problem naturally interfaces with fast methods, such as GPU parallelization or the Fast Multipole Method \citep{Yan2020}, for computing convolutions with  the Laplace Green's function and its derivatives   (see Section \ref{sec:precond_fast_HI}).
Canonical simulations of reactive particles show that the resulting method compares very well with analytical results from the literature with \textit{only a few blobs} on the particle surface  (see Fig. \ref{fig:valid_phoretic}).\\

\itodo{
{\bf Key contributions:}\\
The multiblob method  can simulate large suspensions of particles of arbitrary shape with additional coupling, such as chemical reactions, while accounting for thermal fluctuations.
It employs regularized kernels and a grid optimization strategy to solve the coupled Laplace-Stokes equations with reasonable accuracy at a fraction of the computational cost associated with traditional numerical methods.

The framework is not limited to diffusiophoresis, i.e.\ solute concentration fields. It broadly applies to any scalar field that satisfies elliptic equations, such as temperature or electric potential. With minor modifications, the code can be used to study the dynamics of thermophoretic or electrophoretic particle suspensions. 
}

\section{Applications}
\label{sec:applications}

In this section, we give a brief overview of applications that have been studied with the RMB method, sometimes in close collaboration with experimentalists.
First, we present applications involving suspensions of active particles. These particles are driven either by external magnetic fields, by internal activity, or by hydrochemical interactions. 
Then we present systems involving large suspensions of colloids, where the particle shape and potential interactions are varied to tune the microstructure and rheological response of the suspension.
All of these examples demonstrate the versatility of the RMB to study a wide range of systems in the fields of biophysics, soft and active matter, rheology, materials science and colloids.

\subsubsection{Active particles}
\label{sec:active}

{\bf Microrollers.}
A microroller is a spherical colloid with a permanent magnetic moment that can be rotated by an external magnetic field.
When rotated above a substrate they can \emph{roll} in a direction determined by the external magnetic field. 
Systems formed by microrollers show an interesting interplay between the external drive, hydrodynamic interactions and  thermal fluctuations. 
For example, a dense monolayer of colloids rolling above a flat substrate develops a fingering instability
that eventually seeds motile clusters of colloids which are stable for long periods of time \cite{driscoll2017unstable}. 
The origin of the instability is the strong flows generated by the rotating microrollers,
however, its growth rate and the size of the seeded clusters is controlled by the height of the microrollers above the substrate
\cite{delmotte2017minimal}.
The height, in turn, is affected by the Brownian motion that lifts the colloids above the substrate \cite{Usabiaga2017,Sprinkle2020}.
Therefore, to accurately capture the fingering instability and the emergence of clusters it is necessary to solve the hydrodynamic problem with thermal fluctuations.

\begin{figure} 
  \centering
  \includegraphics[width=0.99 \textwidth]{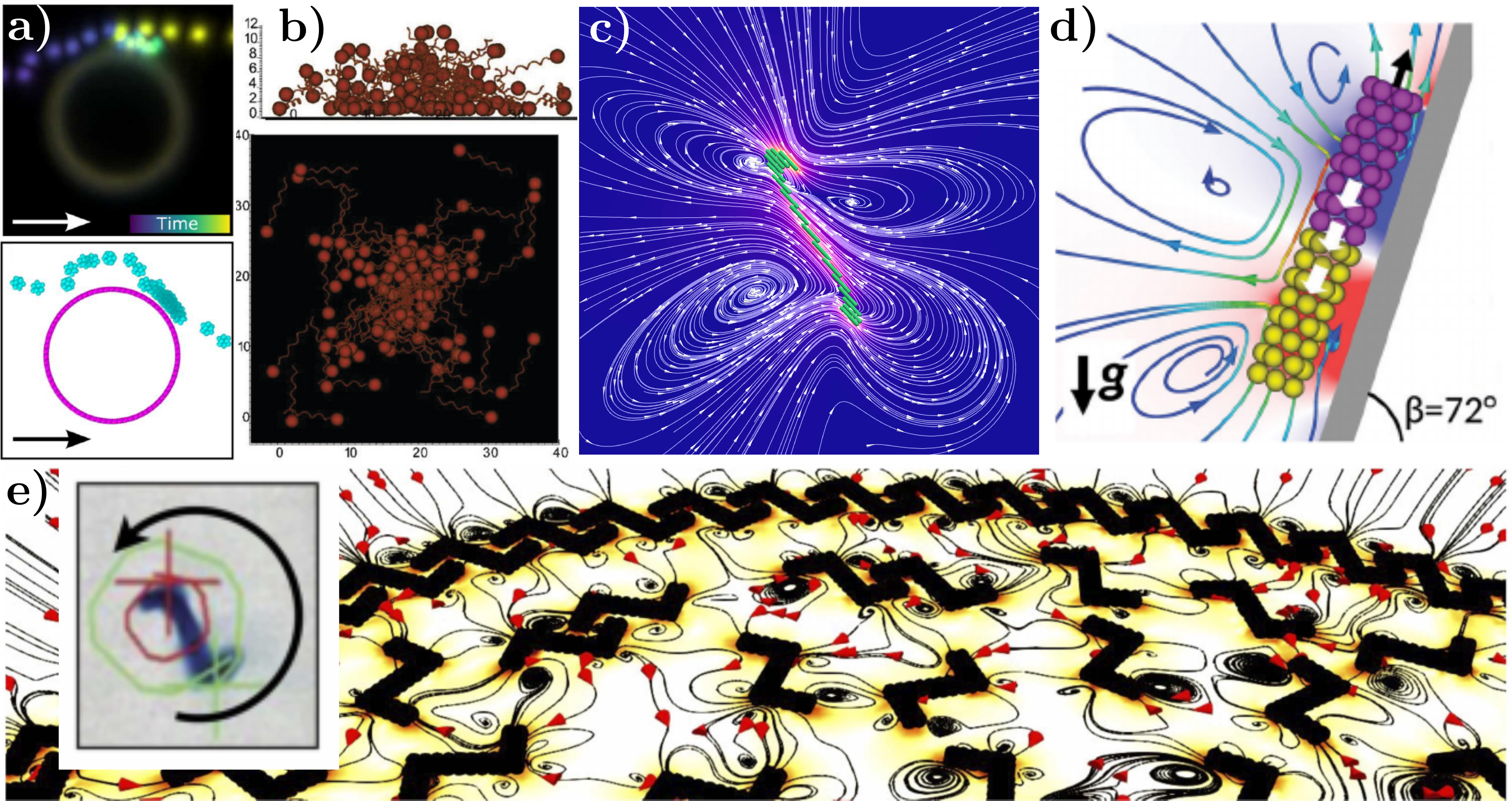}
  \caption{Illustrations of active particle systems simulated with the RMB framework. (a) Superposition of experimental (top) and numerical (bottom) images of a microroller interacting with an obstacle \cite{Van2023}.
    In both cases the microroller remains close to the obstacle for a while before the Brownian motion pushes the microroller away.
    (b) Lateral and top view of 100 bacteria swimming above a no-slip wall \cite{Usabiaga2022}.
    (c) Streamlines around a moving \emph{Bacillaria} colony formed by 16 cells.
    (d) Lateral view of a self-propelled nanorod swimming along a tilted substrate \cite{Brosseau2021}.
    The blobs of different colloids represent two different metals, the experiments used gold-platinum and gold-rhodium nanorods.
    The red (blue) shaded areas show the regions of high (low) pressure while
    the continuous lines represent the streamlines and the white arrows the active slip imposed in the nanorod.
    (e) Flow field around phoretic chiral particles \cite{Delmotte2024}.
    Color-scale from white to dark red: magnitude of the velocity field. Black lines with red arrows: streamlines.
    Inset: experimental reactive stirrer, adapted from \cite{zhang2019reactive}.
  }
  \label{fig:active}
\end{figure}

The interplay between Brownian noise and hydrodynamic interactions affects the dynamics of microrollers in other ways.
For example, a microroller can become trapped by a cylindrical obstacle, with the trapping time increasing with the obstacle size, see Fig.\ \ref{fig:active}a \cite{Van2023}.
The RMB method was used to understand this phenomenon by discretizing both the microroller and the obstacle with a small number of blobs.
and by solving a mixed mobility-resistance problem in which the obstacle was fixed while the microroller was free to move under the action of an external magnetic field.
Thanks to simulations, it was found that the rotation of the microroller creates flows that can be one or two orders of magnitude greater than the self-induced velocity, and that the hydrodynamic response of the obstacle to these flows strongly attracts  the microroller towards the obstacle. The balance between this hydrodynamic attraction and the roller self-induced velocity delimits a basin of attraction that
microrollers can enter or exit only through the diffusion generated by their Brownian motion.
The trapping mechanism is therefore quite unique:  noise, here thermal fluctuations, is the only way to reach an hydrodynamic attractor \cite{Van2023}. 
The basin of attraction size increases with the obstacle size which explains the escape times,
as Brownian displacements are less likely to move a microroller out of a large basin of attraction.
By solving the hydrodynamic and Brownian problems simultaneously, the RMB method successfully reproduced  and explained the experimental results \cite{Van2023}.
\smallbreak

{\bf Microorganisms.} 
The versatility of the RMB method, combined with fast methods to evaluate hydrodynamic interactions (cf.\ Section \ref{sec:precond_fast_HI}),
allows to study a plethora of large constrained systems within a unified framework.
For example, some microorganisms have developed mechanisms that break the kinematic reversibility of Stokes flows to self-propel.
The classical examples are flagellated bacteria (see Fig.\ \ref{fig:context}d) and cilia covered organisms, but there are others, such as diatom  colonies, that move
by their relative displacements \cite{Jahn2007}.
These organisms can be modeled as articulated rigid bodies with the formalism discussed in Secs.\ \ref{sec:constraints} and \ref{sec:articulated},
and the RMB method has been used to study some of them.

For instance, a bacterium can be modeled as a rigid body with a rigid flagellum attached by inextensible links
so that  the flagellum is free to rotate around its main axis but otherwise moves with the bacterium body.
Within this formalism it is possible to apply equal but opposite torques to the bacterium body and flagellum
to mimic the effect of the molecular motor acting on the flagellum, so the whole bacterium swims like a free-force swimmer (see Fig.\ \ref{fig:bacteria_fields}).
This approach was used with the RMB method to model the swimming dynamics of  $100$ bacteria above a no-slip wall, see Fig.\ \ref{fig:active}b \cite{Usabiaga2022}.
The bacteria were hydrodynamically attracted by the no-slip wall \cite{Berke2008} and formed a large cluster,
as shown in the snapshot in Fig.~\ref{fig:active}b, which eventually broke up.
Such a large simulation, with $61,800$ unknowns, 
was possible because the mobility problem was solved in a moderate number of iterations,
because the hydrodynamic interactions were computed with a fast method \citep{Yan2018, Yan2020},
and because it was possible to use large time step sizes as the inextensible links did not introduce any stiffness into the equations of motion.

The articulated body formalism was also used to study the dynamics of \emph{Bacillaria} colonies.
\emph{Bacillaria} is a rod-shaped diatom about $70\si{\micro m}$ long and $10\si{\micro m}$ thick that habits marine and fresh water ecosystems \cite{Jahn2007}.
\emph{Bacillaria} cells can attach to each other along their long axes forming a stack of rods, see Figs.\ \ref{fig:context}c and \ref{fig:active}c.
Once attached, the cells have the ability to slide with respect their neighbors in a coordinated fashion.
In Ref.\ \cite{Usabiaga2022} \emph{Bacillaria} cells were modeled as rods formed by only $14$ blobs and the dynamics of a colony formed by $16$ cells was studied.
The cells were forced to slide against their first neighbors with a sinusoidal pattern and thus,
the whole colony deformed following a wave form. 
Figure \ref{fig:active}c shows a typical snapshot of the colony and the induced flow field. Unlike typical  beating flagella, it was found the swimming direction and speed of the colony varied non-monotonically with the wavelength of the deformation wave \cite{Usabiaga2022}. These nontrivial results are currently investigated and  experiments are carried out in parallel.
\smallbreak

{\bf Phoretic nanorods.}
Another problem, studied with the RMB method, where the interplay between hydrodynamic interactions and Brownian motion is important
is the dynamics of phoretic nanorods.
Bimetallic nanorods can self-propel through electrophoresis in aqueous $\mbox{H}_2\mbox{O}_2$ solutions \cite{Paxton2004}.
These nanorods, similar to the ones shown in Fig.\ \ref{fig:context}g, are formed by two segments of different metals, typically gold and platinum, and decompose the
$\mbox{H}_2\mbox{O}_2$ with asymmetric reactions on each metallic segment \cite{MoranPosner2011}.
The chemical reactions produce a flux of ions near the nanomotor surface that drags the fluid and by momentum conservation propels the nanorod in the opposite direction.
These nanorods have a tendency to swim against flow currents and also against gravity along tilted walls, i.e.\ they show rheotaxis and gravitaxis \cite{Brosseau2019, Brosseau2021}.
RMB simulations were used to understand these behaviors.
In those simulations the nanorods were discretized with a moderated number of blobs ($\sim 100$).
Instead of solving the complex electrochemical reactions an active slip velocity was set \emph{by hand} near the bimetallic interface, see Fig.\ \ref{fig:active}d.

The solution of the hydrodynamic problem showed that the nanorods swim with a tilt towards the substrate, see Fig.\ \ref{fig:active}d.
This tilt makes the nanorods behave like a weather vane and it is the origin of their rheotaxis. 
In the presence of a shear flow above the substrate a tilted rod reorients itself against the flow and once reoriented it tends to swim upstream.
Pure deterministic simulations predict that a nanorod can swim upstream whenever its intrinsic swimming speed is larger than the flow speed
at the height of the nanorod above the substrate.
However, Brownian motion randomizes the nanorod orientation which makes it a less efficient rheotactor. 
RMB simulations were able to quantitatively reproduce the experimental upstream swimming velocities by solving the hydrodynamic and Brownian problems simultaneously \cite{Brosseau2019}.
A similar effect was observed for heavy-bottom nanorods swimming on inclined substrates \cite{Brosseau2021}.
Heavy-bottom nanorods reorient against gravity and swim up showing gravitaxis while Brownian motion randomizes their orientations and thus reduce their gravitactic efficiency.
Interestingly, RMB simulations predicted that density homogeneous rods also experience a reorienting torque.
A nanorod with a tilt towards the substrate rotates around a point displaced towards its head, therefore, even if the center of mass
is at its geometric center the nanorod experiences a reorienting torque.
These numerical findings were confirmed experimentally \cite{Brosseau2021}.
\smallbreak

{\bf Phoretic chiral particles.} 
Thanks to the scalability and versatility of the method, we have been able to explore new physical problems that, to the best of our knowledge,
have not been addressed with numerical simulations. 
Fig.\ \ref{fig:active}e shows an example, inspired by recent experimental work  \citep{Zhang2019,Brooks2019,Kumar2022},
consisting of a large collection of chiral particles, called \emph{reactive stirrers}.
We used the approach described in Secs.\ \ref{sec:Laplace} and \ref{sec:multi-physics}
to model these particles as reacting particles with uniform surface properties that emit solute at a fixed rate \cite{Delmotte2024}.
We confined the particles with a harmonic potential above a no-slip wall.
Thanks to its chirality, a single particle rotates spontaneously due to self-phoresis and stirs the surrounding fluid.
We found that a large collection of such particles exhibit nontrivial collective behaviors,
such as the coexistence of a rotating outer rim
and an inner suspension of particles rotating in opposite directions, see Fig.\ \ref{fig:active}e \cite{Delmotte2024}.

\subsubsection{Complex colloidal suspensions}
\label{sec:suspensions}

\begin{figure} 
  \centering
  \includegraphics[width=0.99 \textwidth]{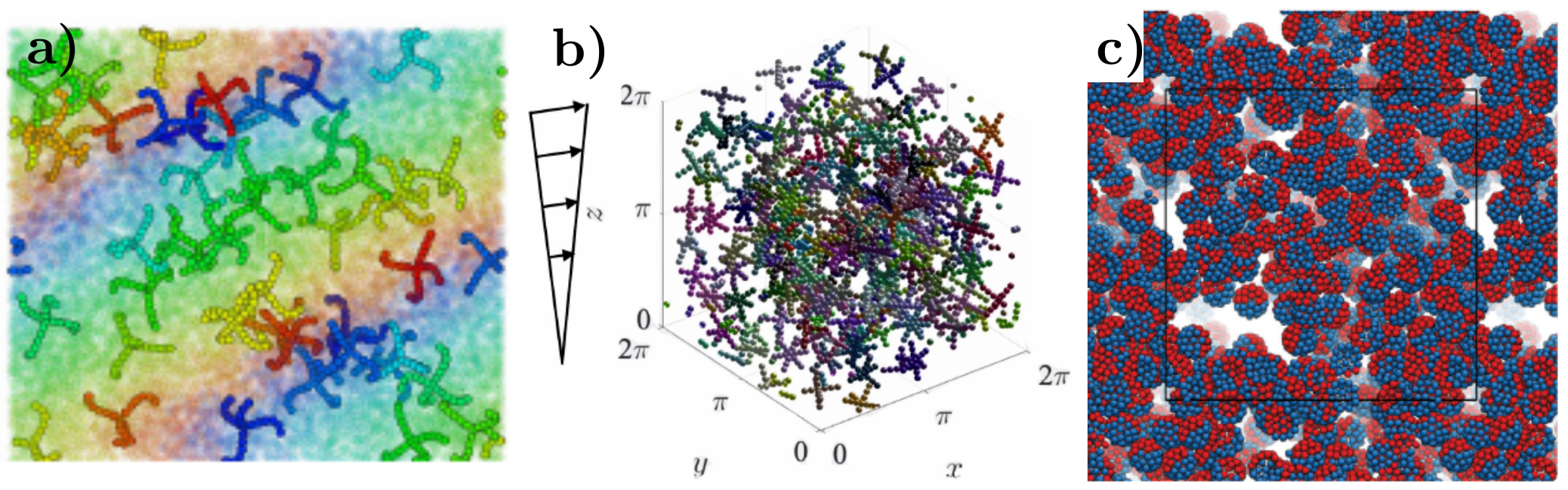}
  \caption{ Examples of complex colloidal suspensions simulated with the RMB framework.
    (a) Snapshot of star shaped colloids under shear flow. A selected number of colloids are shown with full colors
    while the rest are shown with a transparency to easy the visualization. 
    The colloids are colored according to their initial position along the $x$-axis to reveal the shear flow \cite{Balboa2024}.
    (b) Snapshot of the sheared branched particles in a periodic domain. Colors are chosen randomly \cite{Westwood2022}.
    (c) Suspension of patchy colloids with patch depended attraction-repulsion interactions \cite{wang2019surface}.
  }
  \label{fig:suspension}
\end{figure}

{\bf Rheology of complex shaped colloids.}
Some colloids are formed by the 
aggregation of smaller particles.
For example, spherical silica particles can fuse to form complex 
colloids \cite{bourrianne2022tuning}, see Fig.\ \ref{fig:context}b.
Suspensions formed by such colloids have an interesting rheology, exhibiting high viscosities at low volume fractions and continuous or discontinuous shear thickening  \cite{bourrianne2022tuning}.
Modeling such systems with traditional methods is too costly due to the complex shapes of the colloids and the role of Brownian motion in their rheological response.
However, within the RMB's framework each silica particle can be modeled as a single blob which reduces the numerical complexity and cost,
see Fig.\ \ref{fig:suspension}a-b.
RMB studies of star colloids suspensions have shown that their viscosity can be quite large at low and moderate volume fractions in agreement with experimental results.
A hundred fold increase in the viscosity at 20\% volume fraction was observed when the
colloids have complex shapes, like hooks, that allow them to link to each other and form a network that resists stress \cite{Balboa2024}.
However, in contrast with experimental results that found a shear thickening response \cite{bourrianne2022tuning}, RMB studies found shear thinning \cite{Westwood2022, Balboa2024}.
Such difference may come from the lack of friction and lubrication forces in the simulations, although both of them can be incorporated into the RMB formulation \cite{Fiore2019}.

Friction has already been used in the context of Stokesian Dynamics and it can be included as local pairwise force between blobs of different colloids \cite{Mari2014}.
Since friction forces do not modify the linear systems discussed in Sec.\ \ref{sec:RMB} their implementation is simple, although, it may introduce stiffness into the
equations of motion.
Including lubrication corrections modifies the mobility problem \eqref{eq:linear_system}, however, the additional complexity is small.
The idea of including lubrication corrections was first introduced in the Stokesian dynamics method \cite{Durlofsky1987, BradyBossis1988}
and it has been discussed in the context of rigid assemblies of blobs \cite{Sprinkle2020, Fiore2019}.

\smallbreak

{\bf Percolation and gelation of patchy colloids.} 
The same numerical approach discussed in this manuscript was used by Wang \& Swan to study the percolation and gelation of patchy colloids \cite{wang2019surface} (see Fig.\ \ref{fig:context}a for an experimental example).
They studied spherical colloids whose surfaces were covered with two types of patches
exhibiting different attractive and repulsive interactions, see Fig.\ \ref{fig:suspension}c.
Each colloid was discretized with 42 blobs, which were assigned to one of the two patch types.
Using this moderate resolution, they were able to study the dynamics of suspensions consisting of 500 colloids under thermal fluctuations and shear forcing.
They found that, at moderate concentrations, patchy colloids can form microstructures quite different from those of homogeneous particles,
which affect the percolation transition points and the suspensions rheology.

\section{Perspectives}
\label{sec:perspectives}

Thanks to its simplicity and flexibility, the rigid multiblob (RMB) method has been used to simulate a wide variety of systems involving complex suspended particles. However, several challenges remain in generalizing the formulation to cover additional systems of interest and improving its efficiency and accuracy. We discuss some of these challenges here.

The formulation discussed in Sec.\ \ref{sec:governing_HI} can be generalized in several ways. For example, the no-slip boundary conditions can be extended to partial-slip or Navier boundary conditions \cite{Neto2005}.
Navier boundary conditions assume that the flow velocity at solid surfaces does not match the surface velocity but instead \emph{slips} with a velocity proportional to the local velocity gradient.
The proportionality factor between the velocity gradient and the slip velocity is the slip length, which represents the distance across the boundary where the linearly extrapolated flow velocity would vanish.
At nanoscales, even small slip lengths can significantly affect the dynamics of suspended particles \cite{Kamal2021}.
Boundary integral formulations for solving the Stokes equations with Navier boundary conditions exist \cite{Swan2008,Kamal2021,Smith2021},
and the RMB can incorporate them into its formalism \cite{Moreno-Chaparro2025}.
These formulations involve the double-layer operator of the Stokes equations, which is not positive definite. Therefore, the approach used in Sec.\ \ref{sec:therm_fluctu} to generate Brownian velocities is not directly applicable, and how to efficiently generate Brownian velocities in this setting remains an open problem.

In this manuscript, we have only discussed problems where the suspended particles are rigid objects.
However, blobs can also be used to model deformable objects, such as fibers.
Assemblies of rigid and elastic objects can thus be built with blobs without any additional complexity.
Elastic interactions can introduce stiffness into the equations of motion, which may require specialized integrators \cite{Maxian2023}.
This approach has been used, for example, to study the coordinated motion of cilia on a spherical microorganism \cite{westwood2021coordinated}.

The methods used to compute hydrodynamic interactions in Sec.\ \ref{sec:RPY_Lanczos} rely on the Green’s function of the Stokes equations, which requires discretizing only the particle surfaces rather than the full three-dimensional fluid domain. However, these methods are difficult to generalize to flows not governed by elliptic partial differential equations, i.e., flows beyond the Stokes regime, such as viscoelastic flows or flows at finite Reynolds numbers. In such cases, IBM-like methods, as discussed in Sec.\ \ref{sec:FCM}, can be used to solve the flow dynamics. Examples modeling complex particles at finite Reynolds numbers can be found in Refs.\ \cite{Kallemov2016, Sprinkle2019}.

Hydrodynamic interactions between spherical blobs are introduced through the mobility matrix $\bM$, and the efficiency in computing these interactions impacts the overall performance of the RMB. Numerous codes are available to efficiently compute the action of $\bM$ (and of $\bM^{1/2}$ in some cases) \cite{Delong2014,keaveny_fluctuating_2014,Fiore2019,Singh2020,Yan2021,Townsend2024,su2024accelerating,pelaez2025universally,torre2025python}. These codes differ in the domain geometry they support (periodic, bounded, infinite, etc.) and in the programming languages used (C, C++, CUDA, Fortran90, Python, JAX). To be invoked by the multiblob method, these codes must be linked as external libraries. A recent initiative called LibMobility aims to centralize various methods for computing hydrodynamic interactions between blobs in a user-friendly framework. The resulting library is designed for streamlined installation and easy integration into existing codes \cite{fish2024libmobility}. Such efforts are arduous but invaluable to the community and should be encouraged,
even if they remain \textit{poorly rewarded} in terms of scientific recognition.

\begin{figure}
  \centering
  \includegraphics[width= 0.75\textwidth]{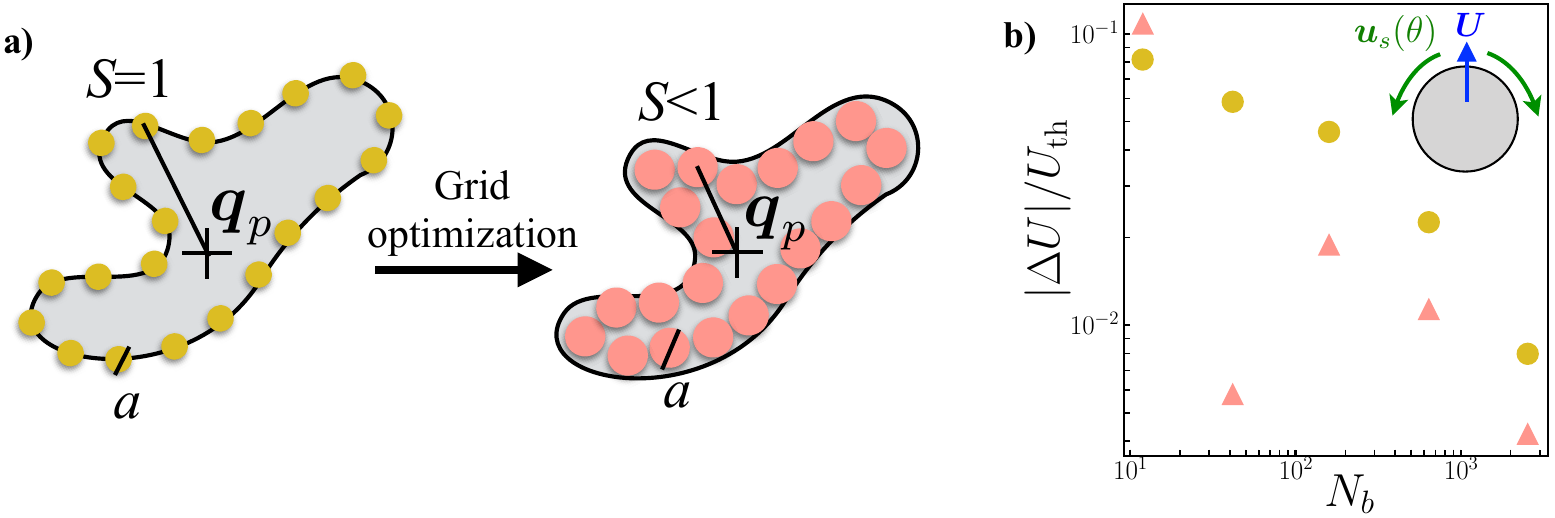}
  \caption{a) Sketch of the grid optimization procedure for a body with arbitrary shape (in gray) and position $\bq_p$. Left: surface discretization of a body with arbitrary shape with $N_b$ blobs of radius $a$ (yellow disks) and size scale $S=1$.  Right: optimized grid with $N_b$ larger blobs (red disks) positioned inside the actual body surface ($S<1$). b)  Relative error of the propulsion speed of a squirmer (sketched in the inset) as a function of the grid resolution $N_b \in [12;2562]$. Disks: empirical  grid. Triangles: optimized grid.
  }
  \label{fig:grid_optim}
\end{figure}

To simulate large numbers of complex particles, it is necessary to use as few blobs per particle as possible while maintaining reasonable accuracy in the hydrodynamic interactions. Despite the regularization error introduced by the regularized Green’s function \eqref{eq:RPY}, which scales with the blob radius $a$ and thus as $N_b^{-1/2}$ \citep{Delmotte2024}, it is possible to enhance numerical accuracy for a given resolution through a grid optimization procedure.
Once the number of blobs per particle and the surface mesh are selected, two additional grid parameters remain to be tuned: the blob radius $a$ and the position of the blobs relative to the actual particle surface.
Due to their finite size, blobs do not need to lie exactly on the particle surface.
The distance of blob centers to the particle surface is described by a geometric parameter $S$, representing the scale of the discretized surface relative to the actual particle surface:
when $S = 1$, the surface of the grid coincides with the surface of the actual particle; if $S < 1$ ($S > 1$), the blobs are inside (outside) the surface (see Fig.\ \ref{fig:grid_optim}a).
In the early works on the multiblob method, this tuning was done empirically and only for the mobility matrix $\bN$ \citep{Usabiaga2016, Swan2016}.
More recently, Broms et al.\ \citep{Broms2023} formalized the optimization of $\bN$ through a minimization problem.
Building on their approach, we proposed a new optimization strategy based on the singular value decomposition of $\wtil{\bN}$ to match the surface mobility operator as well \citep{Delmotte2024}.
Figure \ref{fig:grid_optim}b shows a comparison of the propulsion speed of a spherical squirmer using the empirically chosen grid \cite{Usabiaga2016} and
the optimized grid with the exact solution $U_{\tex{th}}$.
The optimized grid (triangles) significantly reduces the error compared to the original discretization (disks), enabling the simulation of slip-driven particles with coarse grids while achieving 2–3 digits of accuracy—comparable to that of other methods using much finer grids \citep{Smith2021}.
This grid optimization strategy is applicable to particles of any shape. One of the key challenges remaining is the development of quadrature rules that can balance,
or at least further reduce, the regularization error already present in the continuous setting, with the goal of achieving a convergence rate faster than $\sim N_b^{-1/2}$ for the rigid multiblob method. An interesting alternative lies in recent developments of the  method of fundamental solutions, which, although still much more computationally expensive than the RMB, show promising results in terms of accuracy and scalability \cite{broms2025accurate}.

In conclusion, the RMB is a versatile and robust framework for simulating complex particles in Stokes flows. Although there is still room for improvement in its accuracy, efficiency and range of application, it has now reached a stage of maturity that allows it to tackle a variety of physical problems that cannot be addressed using traditional methods.

\section*{Acknowledgments}

The authors would like to pay tribute to Aleks Donev, who was a key contributor to the rigid multiblob method. We also thank the anonymous reviewers for their insightful comments.  B.D.\ acknowledges support from the French National Research Agency (ANR), under award ANR-20-CE30-0006. 
F.B.U.\ acknowledges funding by the Basque Government through the BERC 2022-2025 program and by the Ministry of Science, Innovation and Universities:
BCAM Severo Ochoa accreditation CEX2021-001142-S/MICIN/AEI/10.13039/501100011033.

\bibliography{Biblio}

\end{document}